\documentclass{jfm}

\usepackage{graphicx}
\usepackage{natbib}

\ifCUPmtlplainloaded \else
  \checkfont{eurm10}
  \iffontfound
    \IfFileExists{upmath.sty}
      {\typeout{^^JFound AMS Euler Roman fonts on the system,
                   using the 'upmath' package.^^J}%
       \usepackage{upmath}}
      {\typeout{^^JFound AMS Euler Roman fonts on the system, but you
                   dont seem to have the}%
       \typeout{'upmath' package installed. JFM.cls can take advantage
                 of these fonts,^^Jif you use 'upmath' package.^^J}%
      }
  \else
  \fi
\fi

\ifCUPmtlplainloaded \else
  \checkfont{msam10}
  \iffontfound
    \IfFileExists{amssymb.sty}
      {\typeout{^^JFound AMS Symbol fonts on the system, using the
                'amssymb' package.^^J}%
       \usepackage{amssymb}%
       \let\le=\leqslant  
         
      }{}
  \fi
\fi

\ifCUPmtlplainloaded \else
  \IfFileExists{amsbsy.sty}
    {\typeout{^^JFound the 'amsbsy' package on the system, using it.^^J}%
     \usepackage{amsbsy}}
    {\providecommand\boldsymbol[1]{\mbox{\boldmath $##1$}}}
\fi

\providecommand\bnabla{\boldsymbol{\nabla}}
\providecommand\bcdot{\boldsymbol{\cdot}}

\newsavebox{\astrutbox}
\sbox{\astrutbox}{\rule[-5pt]{0pt}{20pt}}

\newtheorem{remark}{Remark}

\def \FEM {\textbf{FEM~}}
\def \n {\mathbf{n}}
\def \u {\mathbf{u}}
\def \v {\mathbf{v}}
\def \f {\mathbf{f}}
\def \g {\mathbf{g}}

\def \X {\mathbf{X}}
\def \V {\mathbf{V}}
\def \Tau {\mathbf{\tau}}

\title[Spheres in the vicinity of a bifurcation]{Spheres in the vicinity of a bifurcation: \\
elucidating the Zweifach-Fung effect}

\author[V. Doyeux, T. Podgorski, S. Peponas, M. Ismail and G. Coupier]
{V.\ns D\ls O\ls Y\ls E\ls U\ls X$^1$,\ns T.\ns P\ls O\ls D\ls G\ls O\ls R\ls S\ls K\ls I$^1$,\ns \break S.\ns P\ls E\ls P\ls O\ls N\ls A\ls S$^1$,\ns M.\ns I\ls S\ls M\ls A\ls I\ls L$^1$ \and G.\ns C\ls O\ls U\ls P\ls I\ls E\ls R$^1$}

\affiliation{$^1$Laboratoire de Spectrom\'etrie Physique, CNRS et Universit\'e J. Fourier - Grenoble I, BP 87, 38402 Saint-Martin d'H\`eres, France}

\pubyear{}
\volume{}
\pagerange{}
\date{?? and in revised form ??}

\begin{document}

\maketitle

\begin{abstract}

The problem of the splitting of a suspension in bifurcating channels dividing into two branches of non equal flow rates is addressed. As observed for long, in particular in blood flow studies, the volume fraction of particles generally increases in  the high flow rate branch and decreases in the other one. In the literature, this phenomenon is sometimes interpreted as the result of some attraction of the particles towards this high flow rate branch. In this paper, we focus on the existence of such an attraction through microfluidic experiments and two-dimensional simulations and show clearly that such an attraction does not occur but is, on the contrary, directed towards the low flow rate branch. Arguments for this attraction are given and a discussion on the sometimes misleading arguments found in the literature is proposed. Finally, the enrichment in particles in the high flow rate branch is shown to be mainly a consequence of the initial distribution in the inlet branch, which shows necessarily some depletion near the walls.
\end{abstract}


\section{Introduction}
\label{intro}

When a suspension of particles reaches an asymmetric bifurcation, it is well-known that the particle volume fractions in the two daughter branches are not equal; basically, for branches of comparable geometrical characteristics, but receiving different flow rates, the volume fraction of particles increases in the high flow rate branch. This phenomenon, sometimes called the Zweifach-Fung effect \cite[see][]{svanes68,fung73}, has been observed for a long time in the blood circulation. Under standard physiological circumstances, a branch receiving typically one fourth of the blood inflow will see its hematocrit (volume fraction of red blood cells) drop down to zero, which will have obvious physiological consequences. The expression 'attraction towards the high flow rate branch' is sometimes used in the literature as a synonymous for this phenomenon. Indeed, the partitioning not only depends on the interactions between the flow and the particles, which are quite complex in such a peculiar geometry, but also on the initial distribution of particles. \\

Apart the huge number of in-vivo studies on blood flow (see \cite{pries96} for a review), many other papers have been devoted to this effect, either to understand it, or to use it in order to design sorting or purification devices. In the latter case, one can play at will with the different parameters characterizing the bifurcation (widths of the channels, relative angles of the branches), in order to reach a maximum of efficiency. As proposed in many papers, focusing on rigid spheres can already give some keys to understand or control this phenomenon \cite[see][]{bugliarello64,chien85,audet87,ditchfield96,roberts03,roberts06,yang06,barber08}. In-vitro behavior of  red blood cells has also attracted some attention \cite[see][]{dellimore83,fenton85,carr90,yang06,jaggi07,zheng08,fan08}. The problem of particle flow through an array of obstacles, which can be somehow considered as similar, has also been studied recently \cite[see][]{elkareh00,davis06,inglis09,frechette09,balvin09}.

All the latter papers consider the low-Reynolds-number limit, which is the relevant limit for applicative purposes and for the biological systems of interest. Therefore, this limit is also considered throughout this paper.\\

In most studies as well as in in vivo blood flow studies, which are for historical reasons the main sources of data, the main output is the particle volume fraction in the two daughter branches as a function of the flow rate ratio between them. Such data can be well described by empirical laws that still depend on some ad-hoc parameters but allow some rough predictions \cite[see][]{dellimore83,fenton85,pries89}, which have been exhaustively compared recently \cite[see][]{guibert10}.

\begin{figure}
\begin{center}
  \includegraphics[width=\columnwidth]{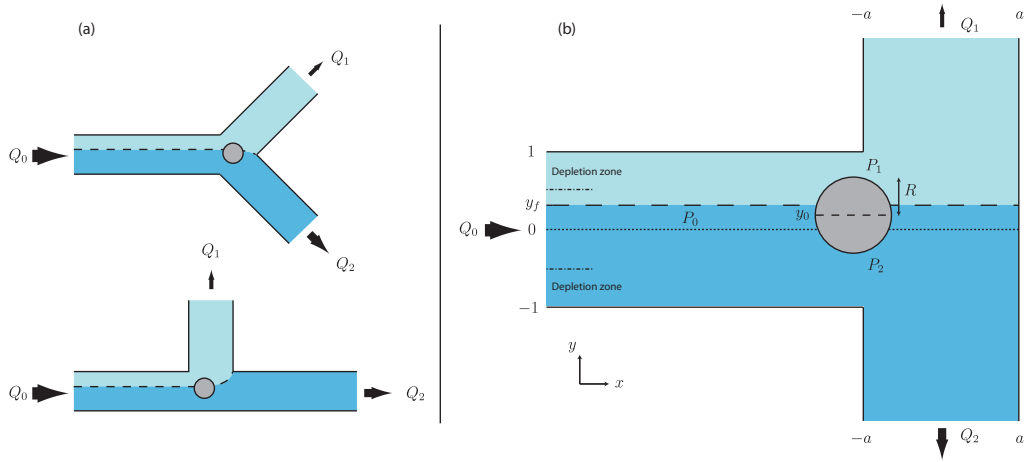}

\caption{(colour online) (a) The two Y-shaped  geometries mainly studied in the literature. Here $Q_1<Q_2$ and the dashed line stands for the separating streamline between the flows that will eventually enter branches 1 and 2 in the absence of particles. (b) The T-bifurcation that is studied in this paper and also in \cite{chien85} in order to get rid of geometrical effects  as much as possible.}\label{fig:schema}
\end{center}
\end{figure}

On the other hand, measuring macroscopic data such as volume fractions does not allow to identify  the relevant parameters and effects involved in this asymmetric partitioning phenomenon.

For a given bifurcation geometry and a given flow rate ratio between the two outlet branches, the final distribution of the particles can be straightforwardly derived from two data: first, their spatial distribution in the inlet; second, their trajectories in the vicinity of the bifurcation, starting from all possible initial positions. If the particles follow their underlying unperturbed streamlines (as would a sphere do in a Stokes flow in a straight channel), their final distribution can be easily computed, although particles near the apex of the bifurcation require some specific treatment, since they cannot approach it as much as their underlying streamline does.

The relevant physical question in this problem is thus to identify the hydrodynamic phenomenon at the bifurcation that would make flowing objects escape from their underlying streamlines, which would have as a consequence that a large particle would be driven towards one branch while a tiny fluid particle located at the same position would go to the other branch.

In order to focus on this phenomenon, we need to identify more precisely the other parameters that influence the partitioning, for a given choice of flow rate ratio between the two branches.\\

\begin{enumerate}
\item \textit{The bifurcation geometry.} \cite{audet87} and \cite{roberts03} made it clear, for instance, that the partitioning in Y-shaped bifurcations depends strongly on the angles between the two branches (see figure \ref{fig:schema}a). For instance, while the velocity is mainly longitudinal, the effective available cross section to enter a perpendicular branch is smaller than in the symmetric Y-shaped case. Even in the latter case, the position of the apex of the bifurcation relatively to the separation line between the fluids going in the two branches might play a role, due to the finite size of the flowing objects.
\item \textit{The radial distribution in the inlet channel.} In an extreme case where all the particles are centered  in the inlet channel and follow the underlying fluid streamline, they all enter in the high flow rate branch; more generally the existence of a particle free layer near the walls favours the high flow rate branch, since the depletion in particles it entails is relatively more important for the low flow rate branch, which receives fluid that occupied less place in the inlet branch. The existence of such a particle free layer near the wall has been observed for long in blood circulation, under the name of plasma skimming. More generally, it can be due to lateral migration towards the centre, which can be of inertial origin (high Reynolds number regime)\cite[see][]{schonberg89,asmolov02,eloot04,kim08,yoo10}, or viscous one. In such a situation of low Reynolds number flow, while a sphere does not migrate transversally due to symmetry and linearity in the Stokes equation, deformable objects such as vesicles (closed lipid membranes) \cite[see][]{coupier08,kaoui09}, red blood cells \cite[see][]{secomb07,bagchi07} that exhibit similar dynamics as vesicles \cite[see][]{abkarian07,vlahovska-cargese}, drops \cite[see][]{mortazavi00,griggs07} or elastic capsules \cite[see][]{secomb07,bagchi07,risso06}, might adopt a shape that allows lateral migration. This migration is due to the presence of walls \cite[see][]{olla97,abkarian02,callens08} as well as to the non-constant shear rate \cite[see][]{kaoui08,danker09}.  Even in the case where no migration occurs, the initial distribution is still not homogeneous: since the barycentre of particles cannot be closer to the wall than their radius, there is always some particle free layer near the walls. This sole effect will favour the high flow rate branch.

\item \textit{Interactions between objects.} As illustrated in \cite{ditchfield96} or \cite{chesnutt09}, interactions between objects tend to smoothen the asymmetry of the distribution, in that the second particle of a couple will tend to go in the other branch as the first one. A related issue is the study of trains of drops or bubbles at a bifurcation, that completely obstruct the channels and whose passage in the bifurcation greatly modifies the pressure distribution in its vicinity, and  thus influences the behaviour of the following element \cite[see][]{engl05,jousse06,schindler08,sessoms09}.\\
\end{enumerate}

In spite of the huge literature on this subject, but probably because of the applicative purpose of most studies, the relative importance of these different parameters are seldom quantitatively discussed, although most authors are fully aware of the different phenomena at stake.

As we want to focus in this paper on the question of cross streamline migration in the vicinity of the bifurcation, we will consider rigid spheres, for which no transverse migration in the upstream channel is expected, that are in the vanishing concentration limit and flow through symmetric bifurcations, that is the symmetric Y-shaped and T-shaped bifurcations shown on figure \ref{fig:schema}, where the two daughter branches have same cross section and are equally distributed relatively to the inlet channel. \\ 

Indeed, this rigid spheres case is already quite unclear in the literature. In the following, we first make a short review of some previous studies that consider a geometrically symmetric situation and thoroughly re-analyze their results in order to detect whether the Zweifach-Fung effect they see is due to initial distribution or to some attraction in the vicinity of the bifurcation, which was generally not done (section \ref{sec:litt}).

We then present in sections \ref{sec:method} and \ref{sec:results} our two-dimensional simulations and quasi-two-dimensional experiments (in a sense that the movement of the three-dimensional objects is planar). We mainly focus on the T-shaped bifurcation, in order to avoid as much as possible the geometrical constraint due to the presence of an apex.

Our main result is that there is some attraction towards the low flow rate branch (section \ref{sec:mig}). This result is then analyzed and explained through basic fluid mechanics arguments, which are compared to the ones previously evoked in the literature. 

In a second time, we discuss which consequences this drift has on the final distribution in the daughter branches. To do so, we focus on what the particles concentrations at the outlets would be in the simplest case, that is particles homogeneously distributed in the inlet channel, with the sole (and unavoidable) constraint that they cannot approach the walls closer than their radius (denominated as \textit{depletion effect} in the following, see figure \ref{fig:schema}(b)). This is done through simulations, which allow us to easily control the initial distribution in particles (section \ref{sec:distr}). Consequences for the potential efficiency of sorting or purification devices are discussed. We finally come back, in section \ref{sec:consistency}, to some of the previous studies found in the literature with which quantitative comparisons can be done in order to check the consistency between them and our results.\\

Before discussing the results from the literature and presenting our own data, we shall introduce useful common notations  (see figure \ref{fig:schema}b).

The half-width of the inlet branch is set as the length scale of the problem. The inlet channel divides into two branches of width $2a$ (the case $a=1$ will be mainly considered here by default, unless otherwise stated), and spheres of radius $R\le 1$ are considered. The flow rate at the inlet is noted $Q_0$, and $Q_1$ and $Q_2$ are the flow rates at the upper and lower outlets ($Q_0=Q_1+Q_2$). In the absence of particles, all the fluid particles situated initially above the line $y=y_f$ will eventually enter branch 1. This line is called the (unperturbed) fluid separating streamline. $y_0$ is the initial transverse position of the considered particle far before it reaches the bifurcation ($| y_0| \le 1-R$). $N_1$ and $N_2$ are the numbers of particles entering branches 1 and 2 by  unit time, while $N_0=N_1+N_2$ have entered the inlet channel. The volume fractions in the branches are $\Phi_i=V N_i/Q_i$, where V is the volume of a particle.

With these notations, we can reformulate our question: if $y_0=y_f$, does the particle experience a net force in the $y$ direction (e. g. a pressure difference) that would push it towards one of the branches, while a fluid particle would remain on the separating streamline (by definition of $y_f$)? If so, for which position $y_0^*$ does this force vanish, so that the particle follows the streamlines and eventually hits the opposite wall and reaches an (unstable) equilibrium position? If $Q_1\le Q_2$ and $y_0^*<y_f$, then one will talk about \textit{attraction towards the low flow rate branch}.

Following these notations, we have:
\begin{eqnarray}
N_1&=&\int_{y_0^*}^1 n(y)u^*_x(y) dy,\label{eq:n1}\\
 Q_1&=&\int_{y_f}^1 u_x(y) dy,\label{eq:q1}
\end{eqnarray}
where $n(y)$ is the mean density in particles at height $y$ in inlet branch, $u_x^*$ and $u_x$ are respectively the particles and flow longitudinal upstream velocities. $N_0$ and $Q_0$ are given by the same formula with $y_0^*=y_f=-1$.

The Zweifach-Fung effect can then be written as follows: if $Q_1/Q_0<1/2$ (branch 1 receives less flow than branch 2) then $N_1/N_0<Q_1/Q_0$ (branch 1 receives even less particles than fluid) or equivalently $\Phi_1<\Phi_0$ (the particle concentration is decreased in the low flow rate branch).\\

\section{Previous results in the literature}
\label{sec:litt}
 
In the literature,  the most common symmetric case that is considered is the Y-shaped bifurcation with daughter branches leaving the bifurcation with a $45^\circ$ angle relatively to the inlet channel, and cross sections identical as the one of the inlet channel (figure \ref{fig:schema}a) \cite[see][]{audet87,ditchfield96,roberts03,roberts06,yang06,barber08}. The T-shaped bifurcation (figure \ref{fig:schema}b) has attracted little attention \cite[see][]{yen78,chien85}.  All studies but \cite{yen78} show results for rigid spherical particles, while some results for deformable particles are  given in \cite{yen78} and \cite{barber08}. Explicit data on a possible attraction towards one branch are scarce as they can only be found in a recent two-dimensional simulations paper \cite[see][]{barber08}. In three other papers, dealing with two-dimensional simulations \cite[see][]{audet87} or experiments in square cross section channels \cite[see][]{roberts06,yang06}, the output data are the concentrations $\Phi_i$ at the outlets. In this section, we re-analyze their data in order to discuss the possibility of an attraction towards one branch. Experiments in circular cross section channels  were also developed \cite[see][]{yen78,chien85,ditchfield96,roberts03}, on which we comment in a second time.

In the two-dimensional simulations presented in \cite{audet87}, some trajectories around the bifurcation are shown, however the authors focused on an asymmetric Y-shaped bifurcation. In addition, some data for $N_1/N_0$ in a symmetric Y-shaped bifurcation and $R=0.5$ are presented. Yang \textit{et al.} ran experiments with  balls of similar size ($R=0.46$) in a symmetric Y-shaped bifurcation with square cross section and also showed data for  $N_1/N_0$ as a function of $Q_1/Q_0$ \cite[see][]{yang06}. Experiments with larger balls ($R=0.8$) in square cross section channels were carried out in \cite{roberts06}. Once again, the output data are the ratios $N_1/N_0$. In both experiments, the authors made the assumption that the initial ball distribution is homogeneous, as considered also in the simulation paper by Audet and Olbricht. In all the latter papers, although the authors are sometimes conscious that the depletion and attraction effects might screen each other, the relative weight of each phenomenon is not really discussed. However, Yang \textit{et al.} consider explicitly that there must be some \textit{attraction towards the high flow rate branch} and give some qualitative arguments for it. This opinion, initially introduced by Fung \cite[see][]{fung73,yen78,fung93}, is widely spread in the literature \cite[see][]{elkareh00,jaggi07,kersaudykerhoas10}. We shall come back to the underlying arguments in the following.\\

\begin{figure*}
\begin{center}

 \includegraphics[width=\columnwidth]{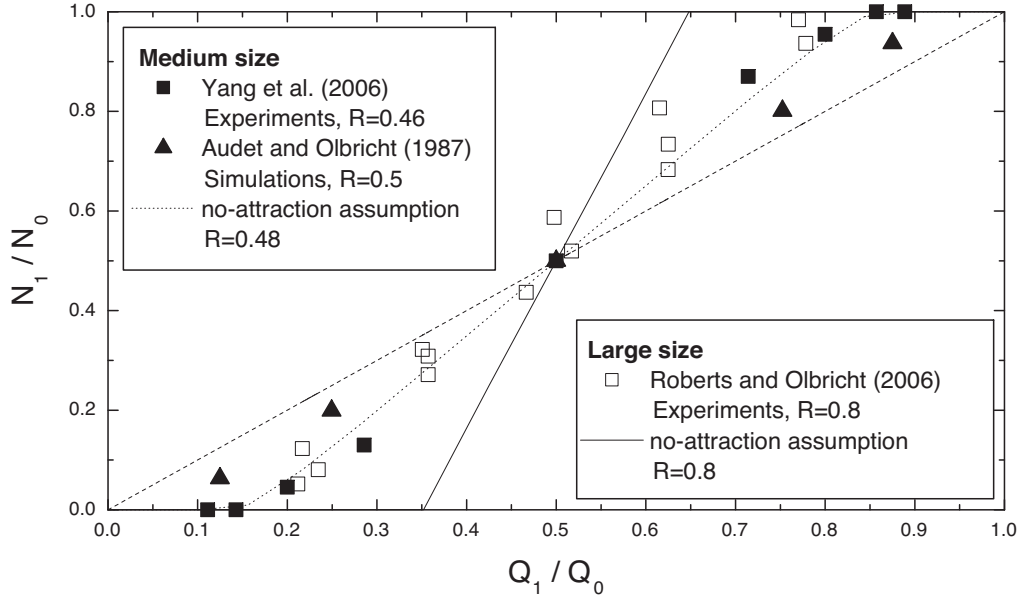}

\caption{Comparison between data from literature and theoretical distribution under the assumption of no attraction, which would indicate some previously unseen attraction towards the low flow rate branch. Rigid spheres distribution $N_1/N_0$ is shown as a function of flow distribution $Q_1/Q_0$ in a symmetric Y-shaped bifurcation and an homogeneous distribution at the inlet (but the unavoidable \textit{depletion effect}). Symbols: data extracted from previous papers: ($\blacksquare$)\cite{yang06},  figure 3, experiments, $R=0.46$. ({\large $\blacktriangle$})\cite{audet87}, figure 8, two-dimensional simulations, $R=0.5$. ($\square$) \cite{roberts06}, figure 5A, experiments, $R=0.8$. Dotted and full lines: theoretical distribution for $R=0.48$ and $R=0.8$ in case the particles follow their underlying streamline ($y_0^*=y_f$: no-attraction assumption) and $u_x^*$ given by our simulations. Dashed line: fluid distribution ($N_1/N_0=Q_1/Q_0$).}\label{fig:compar}
\end{center}
\end{figure*}

In figure \ref{fig:compar} we present the data of $N_1/N_0$ as a function of $Q_1/Q_0$ taken from \cite{audet87} for $R=0.5$ (two-dimensional simulations), \cite{yang06} for $R=0.46$ (experiments) and \cite{roberts06} for $R=0.8$ (experiments). It is very instructive to compare these data with the corresponding values calculated with a very simple model based on the assumption that no particular effect occurs at the bifurcation, that is, the particles follow their underlying streamline (\textit{no-attraction assumption}). To do so, we consider the two-dimensional case of flowing spheres and calculate the corresponding $N_1$ according to equation (\ref{eq:n1}). The no-attraction assumption implies that $y_0^*=y_f$  and, as in the considered papers, the density $n(y)$ is considered constant for $|y| \le 1-R$. The particles velocity $u_x^*$ is given by our simulations presented in section \ref{sec:distr}. Since we consider only flow ratios, this two-dimensional approach is a good enough approximation to discuss the results of the three-dimension experiments, as the fluid separating plane is orthogonal to the plane where the channels lie; moreover, the position of this plane differs only by a few percent from the position of the separating line in two dimensions.

In all curves, it is seen that, if $Q_1/Q_0<1/2$, then $N_1/N_0<Q_1/Q_0$, which is precisely the Zweifach-Fung effect. Note that this effect is present even under the no-attraction assumption: as already discussed, the sole depletion effect is sufficient to favour the high flow rate branch.

Let us first consider spheres of medium size ($R\simeq 0.48 $: Audet and Olbricht / Yang \textit{et al.}). If we compare the data from the literature with the theoretical curve found under the no-attraction assumption, we see that the enrichment in particles in the high flow rate branch is less pronounced in the simulations by Audet and Olbricht and of the same order in the experiments by Yang \textit{et al.}. Therefore, we can assume that in the two-dimensional simulations by Audet and Olbricht, there is an attraction towards the low flow rate branch, which lowers the enrichment of the high flow rate branch. The case of the experiments is less clear: it seems that no peculiar effect takes place.

The $R=0.8$ case is even more striking: under the no-attraction assumption, we can see that for $Q_1/Q_0<0.35$, $N_1=0$ because $y_f>1-R$ and no sphere can enter the low flow rate branch. In the meantime, a non negligible amount of particles are found to enter branch 1 for $Q_1/Q_0<0.35$ by Roberts and Olbricht in their experiments (see figure  \ref{fig:compar}). It is clear from this that there must be some attraction towards the low flow rate branch.\\

For channels with circular cross sections, the data found in the literature do not all tell the same story, although spheres of similar sizes are considered. In \cite{chien85}, $R=0.79$ spheres are considered in a T-shaped bifurcation. The Y-shaped bifurcation was considered twice by the same research group, with very similar spheres: $R=0.8$ \cite[see][]{ditchfield96} and $R=0.77$ \cite[see][]{roberts03}. In a circular cross section channel, the plane orthogonal to the plane where the channels lie, parallel to the streamlines in the inlet channel and located at distance $0.78$ from the inlet channel wall corresponds to the flow separating plane for $Q_1/Q_0=0.32$. At low concentrations, very few spheres are observed in branch 1 for $Q_1/Q_0<0.32$ in \cite{chien85} (figure 3D) and \cite{ditchfield96} (figure 3), in agreement with a no-attraction assumption.  In \cite{chien85}, the authors also show their data can be well described by the theoretical curve calculated by assuming the particles follow their underlying streamlines. In marked contrast with these results, a considerable amount of spheres is still observed in branch 1 in the same situation in \cite{roberts03} (figure 4). Similarly, in \cite{ditchfield96} (figure 4), many particles with $R=0.6$ are found to enter the low flow rate branch 1 even when $Q_1/Q_0<0.19$, which would indicate some attraction towards the low flow rate branch. Thus, in a channel with circular cross section, the results are contradictory. In the pioneering work presented in \cite{yen78}, a T-shaped bifurcation is also considered, with flexible disks mimicking red blood cells, but the deformability of these objects and the noise in the data do not allow us to make any reasonable discussion.\\

More recently, \cite{barber08} presented simulations of two-dimensional spheres with $R\le 0.67$ and two-dimensional deformable objects mimicking red blood cells in a symmetric Y-shaped bifurcation. The values of $y_0^*$ as a function of the flow rate ratios and the spheres radius are clearly discussed. For spheres, it is shown that  $y_0^*<y_f$ if $Q_1<Q_2$, that is, there is an \textit{attraction towards the low flow rate branch}, which increases with $R$. Deformable particles are also considered. However, it is not possible to discuss from their data (as, probably, from any other data) whether the cross streamline migration at the bifurcation is more important in this case or not: for deformable particles, transverse migration towards the centre occurs, due to the presence of walls and of non homogeneous shear rates. This migration will probably screen the attraction effect, at least partly, and it seems difficult to quantify the relative contribution of both effects. In particular, $y_0^*$ depends on the (arbitrary) initial distance from the bifurcation.  In \cite{chesnutt09}, attraction towards the low flow rate branch is also quickly evoked, but considered as negligible since the authors mainly focus on large channels and interacting particles.\\

Finally, from our new analysis of previous results of the literature (and despite some discrepancies) it appears that there should be some attraction towards the low flow rate branch, although the final result is an enrichment of the high flow rate branch due to the depletion effect in the inlet channel. This effect was seen by Barber \textit{et al.} in their simulations. On the other hand, if one considers the flow around an obstacle, as simulated in \cite{elkareh00}, it seems that spherical particles are attracted towards the high flow rate side.\\

From this we conclude that the different effects occurring at the bifurcation level are neither well identified nor explained. Moreover, to date, no direct experimental proof of any attraction phenomenon exists. In section \ref{sec:mig}, we show experimentally that attraction towards the low flow rate branch takes place and confirm this through numerical simulations. 

It is then necessary to discuss whether this attraction has important consequences on the final distributions in particles in the two daughter channels. This was not done explicitly in \cite{barber08}. It is done in section \ref{sec:distr}  where we discuss the relative weight of the attraction towards the low flow rate branch and the depletion effect, which have opposite consequences, by using our simulations.

\section{Method}

\label{sec:method}

\subsection{Experimental setup}
\label{sec:exp}

We studied the behaviour of hard balls as a first reference system. Since the potential migration across streamlines is linked to the way the fluid acts on the particles, we also studied spherical fluid vesicles. They are closed lipid membranes enclosing a Newtonian fluid. The lipids that we used are in liquid phase at room temperature, so that the membrane is a two-dimensional fluid. In particular, it is incompressible (so that spherical vesicles will remain spherical even under stress, unlike drops), but it is easily sheared: it entails that a torque exerted by the fluid on the surface of the particle can imply a different response whether it is a solid ball or a vesicle. Moreover, since vesicle suspensions are polydisperse, it is a convenient way to vary the radius $R$ of the studied object.

\begin{figure}
\begin{center}

  \includegraphics[width=\columnwidth]{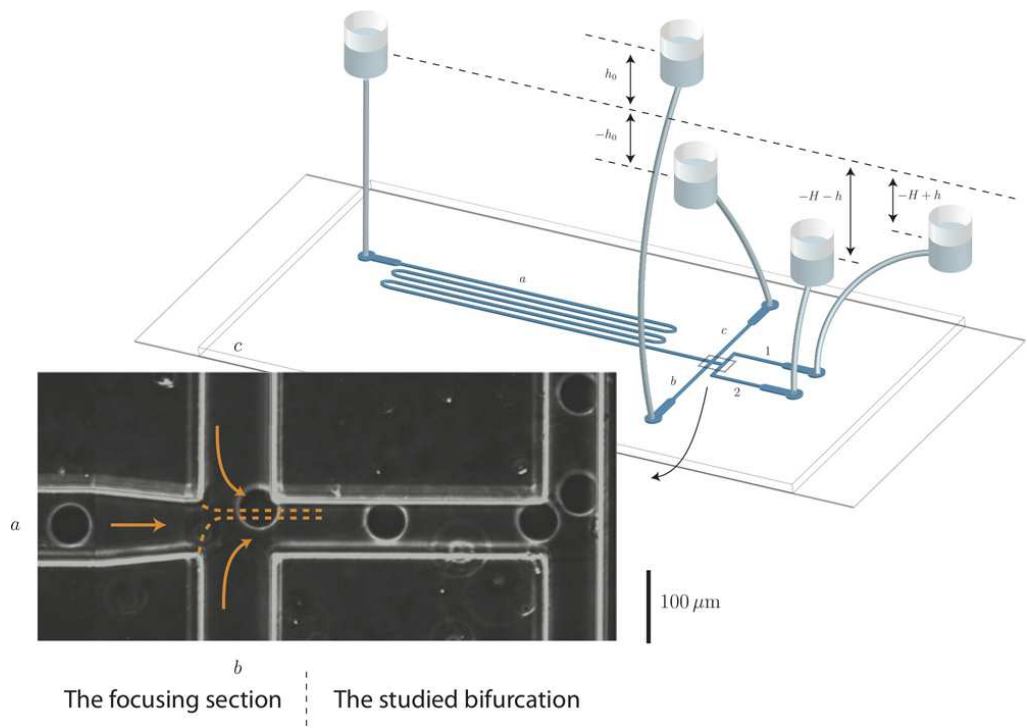}

\caption{(colour online) Scheme of the microfluidic device. The photograph shows the trajectory of a particle from branch $a$ to branch 1 after having been focused on a given streamline thanks to flows from lateral branches $b$ and $c$.}\label{fig:5branches}
\end{center}
\end{figure}

The experimental setup is a standard microfluidic chip made of polydimethylsiloxane bonded on a glass plate (figure \ref{fig:5branches}). We wish to observe what happens to an object located around position $y_f$ that is, in which branch it goes at the bifurcation. In order to determine the corresponding $y_0^*$, we need to scan different initial positions around $y_f$.  One solution would be to let a suspension flow and hope that some of the particles will be close enough to the region of interest. In the meantime, as we shall see, the cross streamline effect is weak and requires precise measurement, and noticeable effects appear only at high radius $R$, typically $R>0.5$. With such objects, clogging is unavoidable, which would modify the flow rates ratio, and if a very dilute  suspension is used, it is likely that the region of interest will only partly be scanned. 

Therefore we designed a microfluidic system allowing to use only one particle, that would go through the bifurcation with a controlled initial position $y_0$, would be taken back, its position $y_0$ modified, would flow again in the bifurcation, and so on. Moreover, we allowed continuous modification of the flow rate ratio between the two daughter branches. The core of the chip is the five branch crossroad shown in inset on figure \ref{fig:5branches}. These five branches have different lengths and are linked to reservoirs placed at different heights, in order to induce a flow by hydrostatic pressure gradient.  A focusing device (branches $a$,$b$ and $c$) is placed before the bifurcation of interest (branches 1 and 2), in order to control the lateral position of the particle. Particles are initially located in the central branch $a$, where the flow is weak and the incoming particles are pinched between the two lateral flows. In order to modify the position $y_0$ of the particle, the relative heights of the reservoir linked to the lateral branches are modified. The total flow rate and the flow rate ratios between the two daughter branches after the bifurcation are controlled by varying the heights of the two outlet reservoirs. Note the flow rates ratio also depends on the heights of the reservoirs linked to inlet branches $a$, $b$ and $c$. Since the two latter must be continuously modified to vary the position $y_0$ of the incoming particle in order to find $y_0^*$ for a given flow rate ratio, it is convenient to place them on a pulley so that their mean height is always constant (the resistances of branches $b$ and $c$ being equal). If the total flow rate is a relevant parameter (which is not be the case here since we consider only Stokes flow of particles that do not deform), one can do the same with the two outlet reservoirs. In such a situation, if reservoir of branch $a$ is placed at height 0, reservoirs of branches $b$ and $c$ at heights $\pm h_0$, and reservoirs of branches 1 and 2 at height $-H+h$ and $-H-h$, the flow rate ratio is governed by setting $(h,H)$ and $h_0$  can be modified independently in order to control $y_0$. Once the particle has gone through the bifurcation, height $H$ and the height of reservoir $a$ are modified so that the particle comes back to branch $a$, and $h_0$ is modified in order to get closer and closer to position $y_0^*$. $Q_1/Q_0$ (or, equivalently, $y_f$) is a function of $H$, $h$, and the flow resistances of the five branches of rectangular cross sections, which are known functions of their lengths, widths and thicknesses \cite[see][]{white}. The accuracy of the calculation of this function was checked by measuring $y_0^*$ for small particles, that must be equal to $y_f$.

Note that the length of the channel is much more important than the size of a single flowing particle, so that we can neglect the contribution of the latter in the resistance to the flow: hence, even though we control the pressures, we can consider that we work at fixed flow rates.

\begin{figure}
\begin{center}

 \includegraphics[width=\columnwidth]{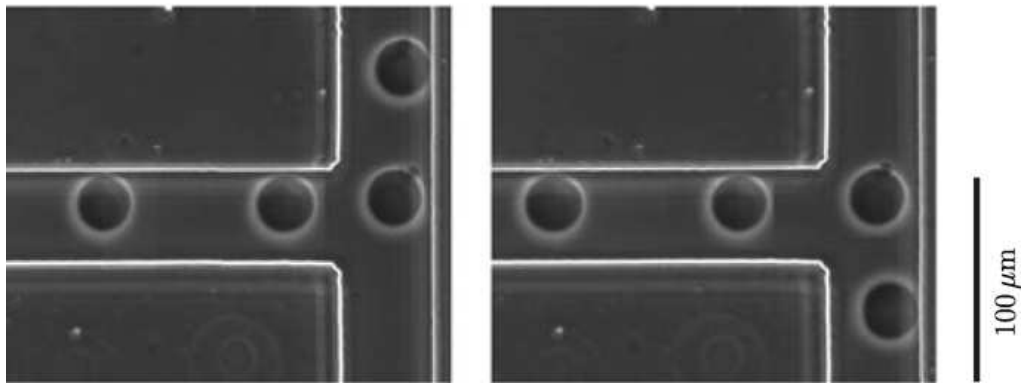}

\caption{Photographs showing the different positions of a vesicle of radius $R=0.60$ starting just above and just below its separating line. No clear difference between these two starting positions can be seen with eyes, which illustrates the accuracy we get in the measurement of $y_0^*$. $Q_1/Q_0$ is set to $0.28$.}\label{fig:discrimination}
\end{center}
\end{figure}

Finally, as it can be seen on figure \ref{fig:discrimination}, our device allows us to scan very precisely the area of interest around the sought $y_0^*$, so that the uncertainty associated to it is very low.\\

At the bifurcation level, channels widths are  all equal to $ 57\pm 0.2 \mu$m. Their thickness is $ 81\pm 0.3 \mu$m. We used  polystyrene balls of maximum radius $40.5 \pm0.3 $ $\mu$m in soapy water (therefore $R\le0.71$) and fluid vesicles of size $R\le 0.60$. Vesicles membrane is a dioleoylphosphatidylcholine lipid bilayer enclosing an inner solution of sugar (sucrose or glucose) in water. Vesicles are produced following the standard electroformation method \cite[see][]{angelova92}. Maximum flow velocity at the bifurcation level was around 1 mm.s$^{-1}$, so that the Reynolds number Re $\simeq 10^{-1}$.

\subsection{The numerical model}

\label{sec:num}

In the simulations, we focus on the two-dimensional problem (invariance along the $z$ axis). Our problem is a simple fluid/structure interaction one and can
be modeled by Navier-Stokes equations for the fluid flow and
Newton-Euler equations for the sphere. These two problems can be
coupled in a simple manner:
\begin{itemize}
\item The action of fluid on the sphere is modeled by the hydrodynamic
  force and torque acting on its surface. They are used as
  the right hand sides of Newton-Euler equations.
\item The action of the sphere on fluid can be modeled by a non-slip
  boundary conditions on the sphere (in the Navier-Stokes equations).
\end{itemize}
However, this explicit coupling can be unstable numerically and 
its resolution often requires very small time steps. In addition, as we have chosen to
use finite element method \FEM (for accuracy reasons) and since the
position of the sphere evolves in time we have to remesh the
computational
domain at each time step or in best cases at each few time steps.

For all these reasons we chose another strategy to model our
problem. Instead of using Newton-Euler equations for modeling the sphere
motion and Navier-Stokes equations for the fluid flow, we use only the
Stokes equations in the whole domain of the bifurcation (including the
interior of the sphere). The use of Stokes equations is justified by the
small Reynolds number in our case and the presence of the sphere is rendered by a second fluid with a 'huge' viscosity on which we impose a
rigid body constraint. This type of strategy is widely used in the
literature with different names e.g. the so called FPD (Fluid Particle
Dynamics) method \cite[see][]{tanaka00,peyla07})
but we can group them under the generic name of penalty-like
methods. The one that we use is mainly developed by Lefebvre {\em et
  al.}  \cite[see][]{janela05,lefebvre07} and we can find a
mathematical analysis of these types of methods in \cite{maury09}.

In what follows we describe briefly the basic ingredients of 
the finite element method and penalty technique applied to our
problem. 

The fluid flow is governed by  Stokes equations that
can be written as follow:
\begin{eqnarray}
  \label{eq:1}
  -\nu\Delta\u + \nabla p &=& 0 \mbox{ in }
  \Omega_f,\\
  \label{eq:2}
  \bnabla\bcdot\u &=& 0 \mbox{ in }
  \Omega_f,\\
  \u &=& \f \mbox{ on } \partial\Omega_f.
  \label{eq:2a}
\end{eqnarray}
Where:
\begin{itemize}
\item $\nu$, $\u$ and $p$ are respectively the viscosity, the velocity
  and the pressure fields of the fluid,
\item $\Omega_f$ is the domain occupied by the fluid. Typically $\Omega_f=\Omega\setminus\bar B$ if we
denote by $\Omega$ the whole bifurcation and by $B$ the rigid
particle,
\item $\partial\Omega_f$ is the border of $\Omega_f$,
\item $\f$ is some given function for the boundary conditions.
\end{itemize}
It is known that under some reasonable assumptions the problem
(\ref{eq:1})-(\ref{eq:2})-(\ref{eq:2a}) has a unique solution
$(\u,p)\in H^1(\Omega_f)^2\times L^2_0(\Omega_f)$ 
\cite[see][]{girault86}. In the sequel we will use the following
functional spaces:
\begin{eqnarray}
  \label{eq:12}
  L^2(\Omega) &=& \{f:\Omega\rightarrow\mathbb{R};
  \int_{\Omega}|f|^2<+\infty\}, \\
  \label{eq:12a}
  L^2_0(\Omega) &=& \{f\in L^2(\Omega); \int_{\Omega}f=0\},\\
  \label{eq:12b}
  H^1(\Omega) &=& \{f\in L^2(\Omega); \nabla f\in L^2(\Omega)\},\\
  \label{eq:12c}
  H^1_0(\Omega) &=& \{f\in H^1(\Omega); f=0 \mbox{ on }\partial\Omega\}.
\end{eqnarray}

As we will use \FEM for the numerical resolution of problem
(\ref{eq:1})-(\ref{eq:2})-(\ref{eq:2a}), we need to rewrite it in a
variational form (an equivalent formulation of the
initial problem).
For sake of simplicity, we start by writing
it in a standard way (fluid without sphere), then we modify it using
penalty technique to take into account the presence of the particle.
In what follow we describe briefly these two methods, the standard variational
formulation for the Stokes problem and the penalty technique.

\subsubsection{Variational formulation}
\label{sec:vari-form}

Let us first recall the deformation tensor $\Tau$ which will be useful in
the sequel 
\begin{equation}
  \label{eq:8}
  \Tau(\u) = \frac{1}{2}\left(\bnabla\u + (\bnabla\u)^t\right).
\end{equation}
Thanks to incompressibility constraint $\bnabla\bcdot\u=0$ we have
\begin{equation}
  \label{eq:11}
  \Delta\u = 2\bnabla\bcdot\Tau(\u).
\end{equation}
Hence, the problem (\ref{eq:1})-(\ref{eq:2})-(\ref{eq:2a}) can be
rewritten as follows: { find 
  $(\u,p)\in H^1(\Omega_f)^2\times L^2_0(\Omega_f)$ such that:}
\begin{eqnarray}
  \label{eq:1b}
  -2\nu\bnabla\bcdot\Tau(\u) + \bnabla p &=& 0 \mbox{ in }
  \Omega_f,\\
  \label{eq:2b}
  \bnabla\bcdot\u &=& 0 \mbox{ in }
  \Omega_f,\\
  \u &=& \f \mbox{ on } \partial\Omega_f.
  \label{eq:2ab}
\end{eqnarray}
By simple calculations (see appendix for details) we
show that problem (\ref{eq:1b})-(\ref{eq:2b})-(\ref{eq:2ab}) is
equivalent to this one:
{ find 
$(\u,p)\in H^1(\Omega_f)^2\times L^2_0(\Omega_f)$ such that:}
\begin{eqnarray}
  \label{eq:3bs}
  2\nu\int_{\Omega_f}\Tau(\u):\Tau(\v)
  -\int_{\Omega_f}p\bnabla\bcdot\v 
  &=& 0, \forall
  \v\in H^1_0(\Omega_f)^2, \\
  \label{eq:4bs}
  \int_{\Omega_f}q\bnabla\bcdot\u &=& 0,
  \forall q\in L^2_0(\Omega_f),\\
  \label{eq:5bs}
  \u &=& \f \mbox{ on } \partial\Omega_f,
\end{eqnarray}

where $:$ denotes the double contraction.
\subsubsection{Penalty method}
\label{sec:penalty-method}

We chose to use the penalty strategy in the framework of \FEM
that we will describe briefly here (see
\cite{janela05,lefebvre07} for more details).

The
first step consists in rewriting the variational formulation
(\ref{eq:3bs})-(\ref{eq:4bs})-(\ref{eq:5bs}) by replacing the integrals
over the real domain occupied by the fluid
($\Omega_f=\Omega\setminus\bar B$) by those over the whole domain
$\Omega$ (including the sphere $B$). Which means that we extend the
solution $(\u,p)$ to the whole domain $\Omega$. More precisely, by the
penalty method we replace the particle by an artificial fluid with
huge viscosity. This is made possible by imposing a rigid body motion
constraint on the fluid that replaces the sphere ($\Tau(\u)=0$ in
$B$). Obviously, the divergence free constraint is also insured in
$B$.

\noindent The problem (\ref{eq:3bs})-(\ref{eq:4bs})-(\ref{eq:4bs}) is then modified
as follows:
{ find 
$(\u,p)\in H^1(\Omega)^2\times L^2_0(\Omega)$ such that:}
\begin{eqnarray}
  \nonumber
  2\nu\int_{\Omega}\Tau(\u):\Tau(\v)
  &+& \frac{2}{\varepsilon}\int_{B}\Tau(\u):\Tau(\v) \\
  \label{eq:3bp}
  &-&\int_{\Omega}p\bnabla\bcdot\v 
  = 0, \forall
  \v\in H^1_0(\Omega)^2, \\
  \label{eq:4bp}
  &&\int_{\Omega}q\bnabla\bcdot\u = 0,
  \forall q\in L^2_0(\Omega),\\
  \label{eq:5bp}
  &&\qquad\quad\;\;\u = \f \mbox{ on } \partial\Omega.
\end{eqnarray}
Where $\varepsilon\ll 1$ is a given penalty parameter.

Finally, if we denote the time discretization parameter by
$t_n=n\delta t$, the velocity and the pressure at time $t_n$ by
$(\u_n,p_n)$, the velocity of the sphere at time $t_n$ by $\V_n$  and its centre position by $\X_n$, we can write our algorithm as:

\begin{eqnarray}
  \label{eq:15}
  \V_n &=& \frac{1}{Volume(B)} \int_B \u_n\\
  \label{eq:15a}
  \X_{n+1} &=& \X_{n} + \delta t\V_{n}
\end{eqnarray}
$(\u_{n+1},p_{n+1})$ solves:
\begin{eqnarray}
  \nonumber
  2\nu\int_{\Omega}\Tau(\u_{n+1}):\Tau(\v)
  &+& \frac{2}{\varepsilon}\int_{B}\Tau(\u_{n+1}):\Tau(\v) \\
  \label{eq:3bpf}
  &-&\int_{\Omega}p_{n+1}\bnabla\bcdot\v 
  = 0, \forall
  \v\in H^1_0(\Omega)^2, \\
  \label{eq:4bpf}
  &&\int_{\Omega}q\bnabla\bcdot\u_{n+1} = 0,
  \forall q\in L^2_0(\Omega),\\
  \label{eq:5bpf}
  &&\qquad\quad\;\;\u_{n+1} = \f \mbox{ on } \partial\Omega.
\end{eqnarray}

The implementation of algorithm 
(\ref{eq:15})-(\ref{eq:15a})-(\ref{eq:3bpf})-(\ref{eq:4bpf})-(\ref{eq:5bpf}) 
is done by using  a user-friendly finite element software: \texttt{Freefem++} 
\cite[see][]{FFpp}.

Finally, we consider the bifurcation geometry shown in figure \ref{fig:schema}(b) and impose no-slip boundary conditions on all walls and we prescribe parabolic velocity profiles at the inlets and outlets such that, for a given choice of flow rate ratio, $Q_0=Q_1+Q_2$. For a given initial position $y_0$ of the sphere of given radius $R$ at the outlet, the full trajectory is calculated until it definitely enters one of the daughter branches. A dichotomy algorithm is used to determine the key position $y_0^*$. Spheres of radius $R$ up to $0.8$ are considered.

\begin{remark}
  In practice, the penalty technique may deteriorate the
  preconditionning of our underlying linear system. To overcome this
  problem, one can regularize equation (\ref{eq:4bpf}) by replacing it
  with this one: 
  \begin{equation}
    \label{eq:16}
    -\varepsilon_0\int_\Omega p_{n+1}q+\int_{\Omega}q\bnabla\bcdot\u_{n+1} = 0,
    \forall q\in L^2_0(\Omega),
  \end{equation}
  where $\varepsilon_0\ll 1$ is a given parameter.
\end{remark}

\section{Results and discussion}
\label{sec:results}

\subsection{The cross streamline migration}
\label{sec:mig}

\begin{figure}
\begin{center}

  \includegraphics[width=0.85\columnwidth]{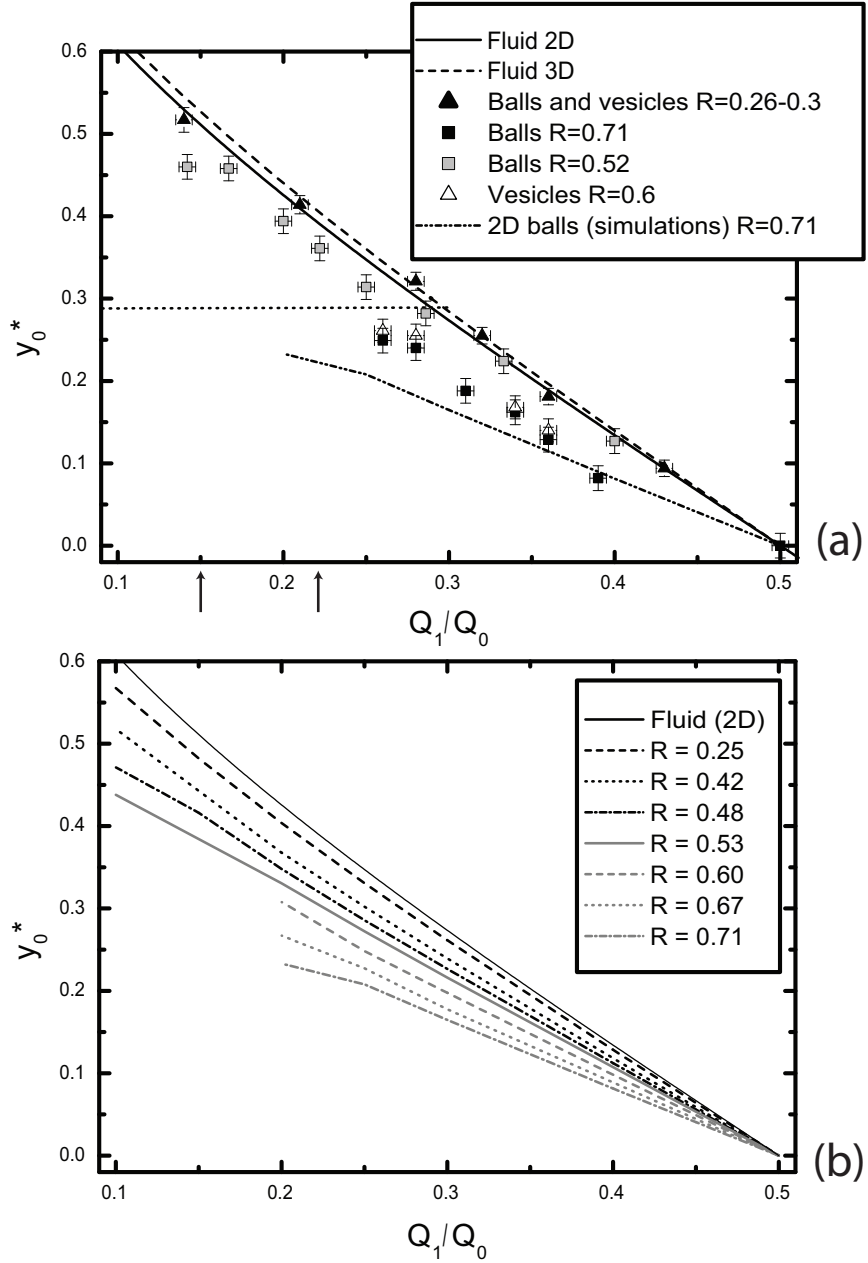}

\caption{Position of the particle separating line $y_0^*$. The T-bifurcation with branches of equal widths is considered. Branch 1 receives flow from high $y$ values, so $y_0^*<y_f$ for $Q_1/Q_0<1/2$ indicates attraction towards the low flow rate branch (see also figure \ref{fig:schema}b). (a) Data from quasi-two-dimensional  experiments and comparison with two-dimensional case for one particle size. The two-dimensional and three-dimensional fluid separating lines are shown to illustrate the low discrepancy between the two cases, as requested to validate our new analysis of the literature in section \ref{sec:litt}. The horizontal dotted line shows the maximum position $y_0=1-R$ for $R=0.71$  spheres. Its intersection with the curve $y_0^*(Q_1/Q_0)$ yields the critical flow rate ratio $Q_1/Q_0$ below which no particle enters branch 1, the low flow rate branch. This expected critical flow rates for the two- and three-dimensional cases are shown by arrows. (b) Data from two-dimensional simulations.}\label{fig:lignessep}
\end{center}
\end{figure}

\subsubsection{The particle separating streamlines}

In figure \ref{fig:lignessep} we show the position of the particle separating line $y_0^*$ relatively to the position of the fluid separating line $y_f$ when branch 1 receives less fluid than branch 2 (see figure \ref{fig:schema}b), which is the main result of this paper. For all particles considered, in the simulations or in the experiments, we find that the particle separating line lies below the fluid separating line, the upper branch being the low flow rate branch. These results clearly indicate an attraction towards the low flow rate branch: while a fluid element located below the fluid separating streamline will enter into the high flow rate branch, a solid particle can cross this streamline and enter into the low flow rate branch, providing it is not too far initially. It is also clear that the attraction increases with the sphere radius $R$.

In particular, in the experiments (figure \ref{fig:lignessep}a), particles of radius $R\lesssim 0.3$ behave like fluid particles. $R=0.52$ balls show a slight attraction towards the low flow rate branch, while the effect is more marked for big balls of radius $R=0.71$. Vesicles show comparable trend and it seems from our data that solid particles or vesicles with fluid membrane behave similarly in the vicinity of the bifurcation.

In the simulations (figure \ref{fig:lignessep}b) we see clearly that for a given $R$, the discrepancy between the fluid and particle behaviour increases when $Q_1/Q_0$ decreases. On the contrary, in the quasi-two-dimensional case of the experiments, the difference between the flow and the particle streamlines seems to be rather constant in a wide range of $Q_1/Q_0$ values. Finally, for small enough values of $Q_1/Q_0$, the attraction effect is more pronounced in the two-dimensional case than in the quasi-two-dimensional one, as shown on figure \ref{fig:lignessep}(a) for $R=0.71$. This was to be expected, since this effect has something to do with the non zero size of the particle and the real particle to channel size ratio is lower in the experiments for a given $R$, due to the third dimension. In all cases, below a given value of $Q_1/Q_0$, the critical position $y_0^*$ would enter the depletion zone $y_0>1-R$, so that no particle will eventually enter the low flow rate branch. The corresponding critical  $Q_1/Q_0$ is much lower in the two-dimensional case than in the experimental quasi-two-dimensional situation (see figure \ref{fig:lignessep}a).

\subsubsection{Discussion}
\label{sec:discussion}

The first argument for some attraction towards one branch was initially given by Fung \cite[see][]{fung73,yen78,fung93} and strengthened by recent simulations  \cite[see][]{yang04}: a sphere in the middle of the bifurcation is considered ($y_0=0$) and it is argued that it should go to the high flow rate branch since the pressure drop $P_0-P_2$ is higher than $P_0-P_1$ because $Q_2>Q_1$ (see figure \ref{fig:schema}(b) for notations). This is true (we also found $y_0^*>0$ when $Q_1<Q_2$) but this is not the point to be discussed: if one wishes to discuss the increase in volume fraction in branch 2, therefore to compare the particles and fluid fluxes $N_2$ and $Q_2$, one needs to focus on particles in the vicinity of the fluid separating streamline (to see whether or not they behave like the fluid) and not in the vicinity of the middle of the channel. On the other hand, this incorrectly formulated argument by Fung has led to the idea that there must be some attraction towards the high flow rate branch in the vicinity of the fluid separating streamline \cite[see][]{yang06}, which appears now in the literature as a well established fact \cite[see][]{jaggi07,kersaudykerhoas10}.

In \cite{barber08}, Fung's argument is rejected, although it is not explained why. Arguments for attraction towards the low flow rate branch (that is, $P_2>P_1$ on figure \ref{fig:schema}(b)) are given, considering particles in the vicinity of the fluid separating streamline. The authors' main idea is, first, that some pressure difference $P_0-P_i$ builds up on each side of the particle because it goes more slowly than the fluid. Then, as the particle intercepts a relatively more important area in the low flow rate branch region ($y_f<y<1$) than in the high flow rate region, they consider that the pressure drop is more important in the low flow rate region, so that $P_2>P_1$. The authors call this effect 'daughter vessel obstruction'.

Indeed, it is not clear in this paper where the particles must be for this argument to be valid: at the entrance of the bifurcation, in the middle of it, or close to the opposite wall as we could think since their arguments are used to explain what happens in case of daughter branches of different widths. Indeed, we shall see that the effects can be  quite different according to this position and, furthermore, the notion of 'relatively larger part intercepted' is not the key phenomenon to understand the final attraction towards the low flow rate branch, even though it clearly contributes to it.

\begin{figure}
\begin{center}

  \includegraphics[width=\columnwidth]{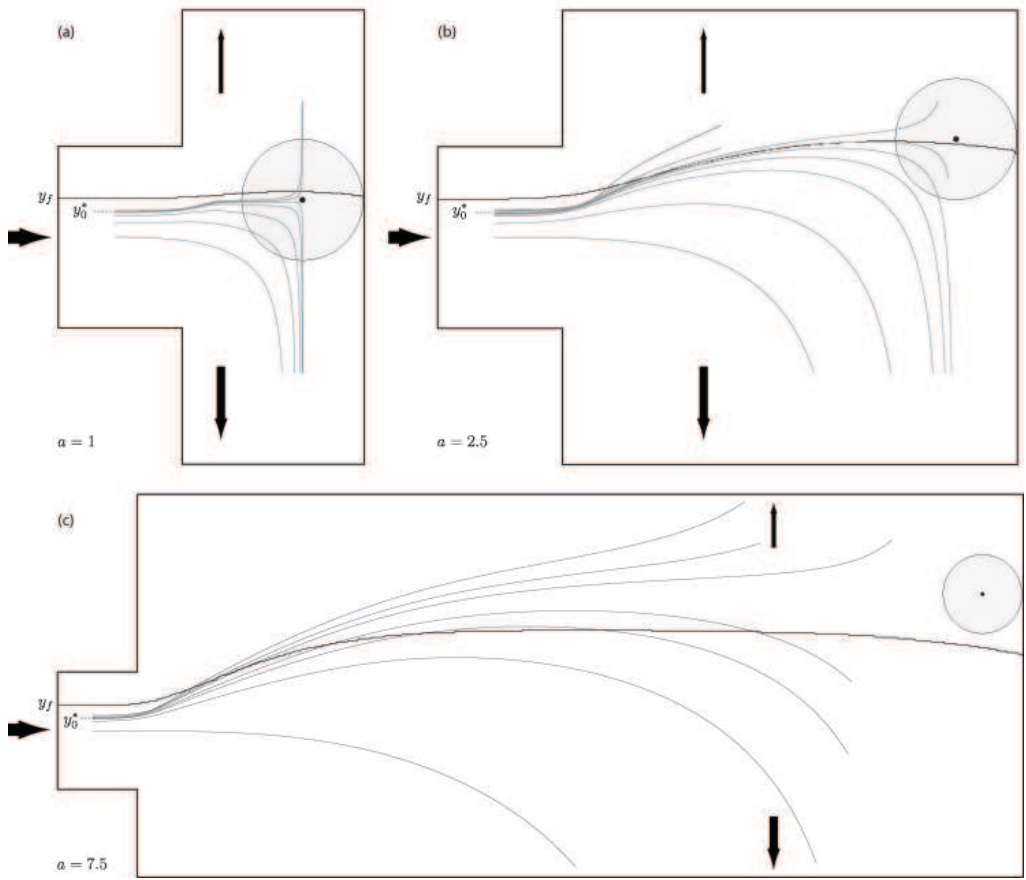}

\caption{Grey lines: some trajectories of a $R=0.67$ particle  when $Q_1/Q_0=0.2$  for (a) branches of equal widths, (b) daughter branches 2.5 times wider than the inlet branch and (c) daughter branches 7.5 times wider than the inlet branch. The unperturbed fluid separating streamline starting at $y=y_f$ is shown in black. The particle is shown approximatively at its stagnation point.}\label{fig:traj}
\end{center}
\end{figure}

To understand this, let us focus on the simulated trajectories starting around $y_0^*$ shown on figure \ref{fig:traj}(a) ($R=0.67$, $Q_1/Q_0=0.2$). These trajectories must be analysed in comparison with the unperturbed flow streamlines, in particular the fluid separating streamline, starting at $y=y_f$ and ending up against the front wall at a stagnation point.

Particles starting around $y_0^*<y_f$ show a clear attraction towards the low flow rate branch (displacement along the $y$ axis) as they enter the bifurcation. More precisely, there are three types of motions: for low initial position $y_0$ (in particular $y_0=0$), particles go directly into the high flow rate branch. Similarly, above $y_0^*$, the particles go directly into the low flow rate branch. Between some $y_0^{**}>0$ and $y_0^*$, the particles first move towards the low flow rate branch, but finally enter the high flow rate branch: the initial attraction towards the low flow rate branch becomes weaker and the particle eventually follows the streamlines entering the high flow rate branch. This non monotonous variation of $y_0$ for a particle starting just below $y_0^*$ is also seen in experiments, as shown in figure \ref{fig:discrimination}, right part: the third position of the vesicle is characterized by a $y_0$ slightly higher than the initial one. Back to the simulations, note that, at this level, there is still some net attraction towards the low flow rate branch: the particle stagnation point near the opposite wall is still below the fluid separating streamline (that is, on the high flow rate side). This two-step effect is even more visible when the width $2a$ of the daughter branches is increased, so that the entrance of the bifurcation is far from the opposite wall, as shown on figures \ref{fig:traj}(b,c). The second attraction is, in such a situation, more dramatic: for $a=7.5$, the particle stagnation point is even on the other side of the fluid separating streamline, that is, there is some attraction towards the high flow rate branch! Thus, there are clearly two antagonistic effects along the trajectory. In the first case of branches of equal widths,  where the opposite wall is close to the bifurcation entrance, the second attraction towards the high flow rate branch coexists with the attraction towards the low flow rate branch and finally only diminishes it. \\

These two effects occur in two very different situations. At the entrance of the channel, an attraction effect must be understood in terms of streamlines crossing: does a pressure difference build up orthogonally to the main flow direction? Near the opposite wall, the flow is directed towards the branches and being attracted means flowing up- or downstream. In both cases, in order to discuss whether some pressure difference builds up or not, the main feature is that, in a two-dimensional Stokes flow between two parallel walls, the pressure difference between two points along the flow direction scales like  $\Delta P \propto Q/h^3$, where $Q$ is the flow rate and $h$ the distance between the two walls. This scaling is sufficient to discuss in a first order approach the two effects at stake.\\

\begin{figure}

  \includegraphics[width=\columnwidth]{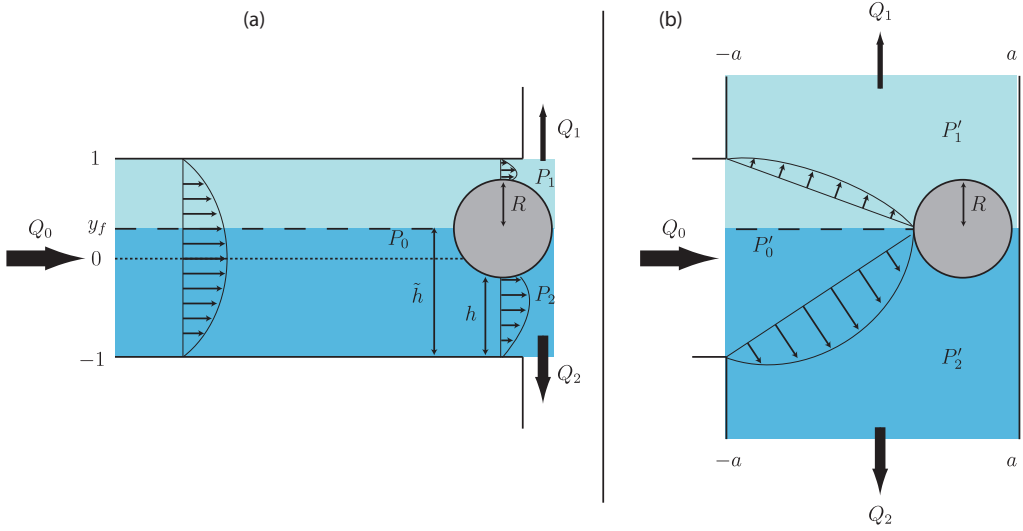}

\caption{(colour online) Scheme of the geometry considered for the two effects occurring in the bifurcation. (a) Entrance of the bifurcation: attraction towards the low flow rate branch ($P_1<P_2$). (b) opposite wall: attraction towards the high flow rate branch ($P'_1>P'_2$).}\label{fig:schemasimple}
\end{figure}

The second effect is the simplest one: indeed, the sphere is placed in a quasi-elongational, but asymmetric, flow. As shown on figure \ref{fig:schemasimple}(b), around the flow stagnation point, the particle movement is basically controlled by the pressure difference $P_2'-P_1'$, than can be written $(P_0'-P_1')-(P_0'-P_2')$. Focusing on the $y$ component of the velocity field, which becomes all the more important as $a$ is larger than 1, we have $P_0'-P_i' \propto Q_i/(a-R)^3$.  Around the flow stagnation point, the pressure difference $P_2'-P_1'$  has then the same sign as $Q_1-Q_2$ and is thus negative, which indicates attraction towards the high flow rate branch. For wide daughter branches, when this effect is not screened by the first one, this implies that the stagnation point for particles is above the fluid separating line, as seen on figure \ref{fig:traj}(c). The argument that we use here is similar to the one introduced by Fung \cite[see][]{fung93,yang06} but resolves only one part of the problem. Following these authors, it can also be pointed out that the shear stress on the sphere is non zero: in a two-dimensional Poiseuille flow of width $h$, the shear rate near a wall scales as $Q/h^2$, so the net shear stress on the sphere is directed towards the high flow rate branch, making the sphere roll along the opposite wall towards this branch.
 
Finally, this situation is similar to the one of a flow around an obstacle, that was considered in \cite{elkareh00} as a model situation to understand what happens at the bifurcation. Indeed, the authors find that spheres are attracted towards the high velocity side of the obstacle. However, we show here that this modeling is misleading, as it neglects the first effect, which is the one which eventually governs the net effect.\\

This first effect leads to an attraction towards the low flow rate branch. To understand this, let us consider a sphere located in the bifurcation with transverse position $y_0=y_f$. The exact calculation of the flow around it is much too complicated, and simplifications are needed. Just as we considered the large $a$ case to understand the second mechanism eventually leading to attraction towards the high flow rate branch, let us consider the small $a$ limit to understand the first effect: as soon as the ball enters the bifurcation, it hits the front wall. On each side, we can write in a first approximation that the flow rate between the sphere and the wall scales as  $Q\propto \Delta P h^3$, where $\Delta P=P_0-P_i$ is the pressure difference between the back and the front of the sphere, and $h$ the distance between the sphere and the wall (see figure \ref{fig:schemasimple}a). 

Since the ball touches the front wall, the flow rate $Q$ is either $Q_1$ or $Q_2$ and is, by definition of $y_f$, the integral of the unperturbed Poiseuille flow velocity between the wall and the $y=y_f$ line, so $Q \propto \tilde{h}^2 -\tilde{h}^3/3$, where $\tilde{h}=1 \pm y_f$ (see figure \ref{fig:schemasimple}(a) for notations).
 
We have then, on each side:
\begin{equation}
\Delta P \propto \frac{\tilde{h}^2 -\tilde{h}^3/3}{h^3} . \label{eq:total}\end{equation}

To make the things clear, let us consider then the extreme case of a flat particle: $h=\tilde{h}$. Then $\Delta P\propto 1/\tilde{h}-1/3$ is a decreasing function of $\tilde{h}$, that is, a decreasing function of $Q$. Therefore, the pressure drop is more important on the low flow rate side, and finally $P_1<P_2$: there is an attraction towards the low flow rate branch. This is exactly the opposite result from the simple view claiming that there is some attraction towards the high flow rate branch since $\Delta P$ scales as $Q/h^3$ so as $Q$. Since one has to discuss what happens for a sphere in the vicinity of the separating line, $Q$ and $\tilde{h}$ are not independent. This is the key argument. Note finally that there is no need for some obstruction arguments to build up a different pressure difference on each side. It only increases the effect since the function $\tilde{h}\mapsto (\tilde{h}^2 -\tilde{h}^3/3)/(\tilde{h}-R)^3$ decreases faster than the function $\tilde{h}\mapsto (\tilde{h}^2 -\tilde{h}^3/3)/\tilde{h}^3$.  One can be even more precise and take into account the variations in the gap thickness as the fluid flows between the sphere to calculate the pressure drop by lubrication theory. Still, it is found that $\Delta P$ is a decreasing function of $\tilde{h}$.\\

In the more realistic case $a\simeq 1$, the flow repartition becomes more complex, and the particle velocity along the $x$ axis is not zero. Yet, as it is reaching a low velocity area (the velocity along $x$ axis of the streamline starting at $y_f$ drops to 0), its velocity is lower than its velocity at the same position in a straight channel. In addition, as the flow velocities between the sphere and the opposite wall are low, and since the fluid located e. g. between $y_f$ and the top wall will eventually enter the top branch by definition, we can assume it will mainly flow between the sphere and the top wall. Note this is not true in a straight channel: there are no reasons for the fluid located between one wall and the $y=y_0$ line, where $y_0$ is the sphere lateral position, to enter completely, or to be the only fluid to enter, between the wall and the particle. Therefore, we can assume that the arguments proposed to explain the attraction towards the low flow rate branch remain valid, even though the net effect will be weaker.

Note finally that, contrary to what discussed for the second effect, the particle rotation probably plays a minor role here, as in this geometry the shear stress exerted by the fluid on the particle will mainly result  in a force acting parallel to the $x$ axis.\\

Finally, this separation into two effects can be used to discuss a scenario for bifurcations with channels of different widths: if the inlet channel is broadened, the first effect becomes less strong while the second one is not modified, which results in a weaker attraction towards the low flow rate branch. If the outlet channels are broadened, as in figures \ref{fig:traj}(b,c), it becomes more subtle. Let us start again by the second effect (migration up- or downstream) before the first effect (transverse migration). As seen on figure \ref{fig:traj}, the position of the particle stagnation point (relatively to the flow separating line) is an increasing function of $a$, so the second effect is favoured by the broadening of the outlets: for $a \to \infty$, we end up with the problem of flow around an obstacle, while for small $a$, one cannot write that the width of the gap between the ball and the wall is just $a-R$, therefore independent from $Q_i$, as it also depends on the $y$ position of the particle relatively to $y_0'$. In other words, in such a situation, the second effect is screened by the first effect. On the other hand, as $a$ increases, the distance available for transverse migration becomes larger, which could favour the first effect,  although the slow down of the particle at the entrance of the bifurcation becomes less pronounced. 

Finally, it appears to be difficult to predict the consequences of an outlet broadening: for instance, in our two-dimensional simulations presented in figure \ref{fig:traj} ($R=0.67$, $Q_1/Q_0=0.2$), $y_0^*$ varies from 0.27 when the outlet half-width $a$ is equal to 1, to 0.31 when $a$ is equal to 2.5 and drops down to 0.22 for $a=7.5$! Note that the net effect is always an attraction towards the low flow rate branch ($y_0^*<y_f$).

For daughter branches of different widths, it was illustrated in \cite{barber08} that the narrower branch is favoured. This can be explained through the second effect (see figure \ref{fig:schemasimple}b): the pressure drop $P_0'-P_i' \propto Q_i/(a-R)^3$ increases when the channel width decreases, which favours the narrower branch even in case of equal flow rates between the branches.

\subsection{The consequences on the final distribution}
\label{sec:distr}

\begin{figure}

  \includegraphics[width=\columnwidth]{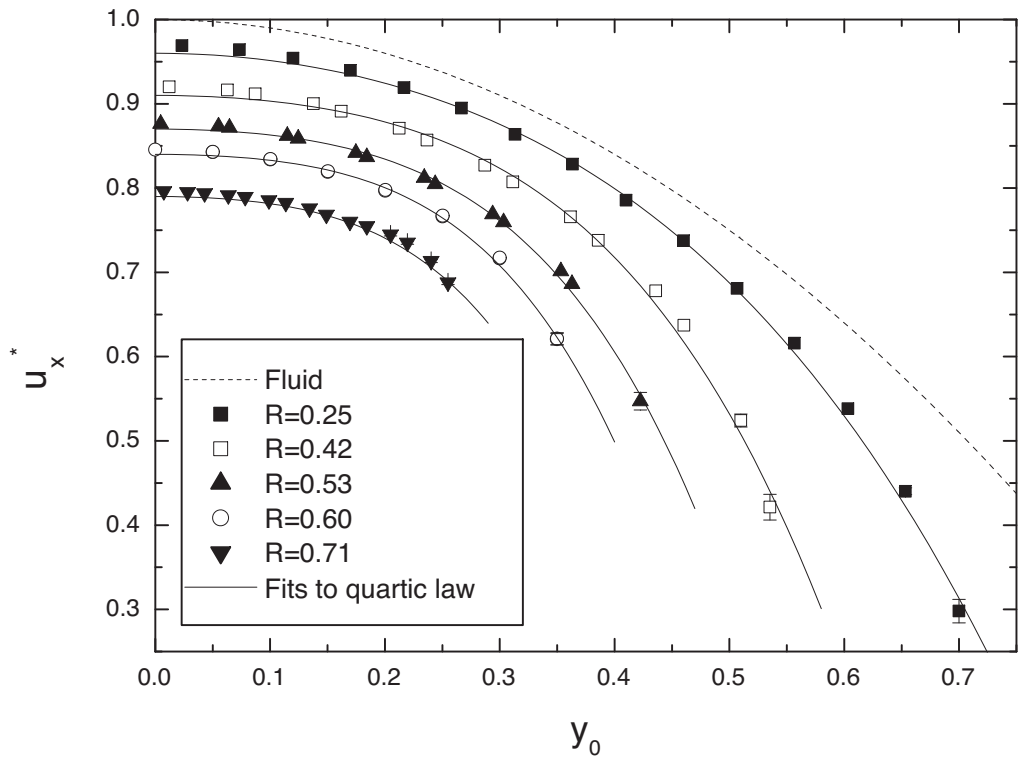}

\caption{Some longitudinal velocity profiles $u_x^*(y_0)$ for two-dimensional spheres of different radii $R$ in the inlet channel where a Poiseuille flow of velocity $u_x(y)=1-y^2$ is imposed at infinity. The full lines show the fits by quartic law $u_x^*(y_0)=\alpha y_0^4+\beta y_0^2 + \gamma$.}\label{fig:profils}
\end{figure}

As there is some attraction towards the low flow rate branch, we could expect some enrichment of the low flow rate branch. However, as already discussed, even in the most uniform situation, the presence of a free layer near the walls will favour the high flow rate branch. We discuss now, through our simulations, the final distribution that results from these two antagonistic effects.

As in most previous papers of the literature, we focus on the case of uniform number density of particles in the inlet ($n(y)=1$ in equation \ref{eq:n1}). In order to compute the final splitting $N_1/N_0$ of the incoming particles as a function of flow rate ratio $Q_1/Q_0$ one needs to know, according to equation (\ref{eq:n1}), the position $y_0^*$ of the particle separating line and the velocity $u_x^*$ of the particles in the inlet channel. From figure \ref{fig:lignessep} we see that $y_0^*$ depends roughly linearly on $(Q_1/Q_0-1/2)$, so we will consider a linear fit of the calculated data in order to get values for all $Q_1/Q_0$. The longitudinal velocity $u_x^*$ was computed for all studied particles as a function of transverse position $y_0$. As shown on figure \ref{fig:profils}, the function $u_x^*(y_0)$ is well described by a quartic function $u_x^*(y_0)=\alpha y_0^4+\beta y_0^2 + \gamma$, which is an approximation also used in \cite{barber08}. Values for the fitting parameters for this velocity profile and for the linear relationship $y_0^*= \xi \times(Q_1/Q_0-1/2)$ are given in table \ref{table:quartic}.

\begin{table}
  \begin{center}

  \begin{tabular}{lccccccccc}
  
  $R$ & 0 & 0.25            & 0.42    & 0.48     & 0.53 & 0.60   & 0.67   &0.71 &0.80 \\
  $\alpha$ & 0 & -0.96    & -3.45  &   -4.77   & -6.33 & -9.52 & -11.6  & -12.6 & $-$\\
  $\beta$ & -1 &  -0.85   & -0.65  &     -0.70  & -0.64  & -0.61  & -0.71  &-0.73&$-$ \\
  $\gamma$ & 1 & 0.96  & 0.91  &     0.89   & 0.87 & 0.84    & 0.81   &0.79&0.75\\
  $\xi$&$-$&-1.35&-1.25&-1.17&-1.09&-1.01&-0.90&-0.81&$-$\\ 
  \end{tabular}
  \caption{Values for the fitting parameters $(\alpha,\beta,\gamma)$ for the longitudinal velocity $u_x^*(y_0)=\alpha y_0^4+\beta y_0^2 + \gamma$ of a two-dimensional sphere of radius $R$ in a Poiseuille flow of imposed velocity at infinity $u_x(y)=1-y^2$; for $R=0.80$, the velocity profile is too flat to be reasonably fitted by a 3-parameter law, since all velocities are equal to $0.75\pm0.005$ in the explored interval $y_0\in [-0.15;0.15]$. We also give the values for the fitting parameter $\xi$ of the linear relationship between particle separating line position $y_0^*$ and flow rate ratios: $y_0^*= \xi \times(Q_1/Q_0-1/2)$. For $R=0.8$, the strong confinement leads to numerical problems as the sphere approaches the walls.}
  \label{table:quartic}
  \end{center}
\end{table}

\begin{figure}
\begin{center}

  \includegraphics[width=\columnwidth]{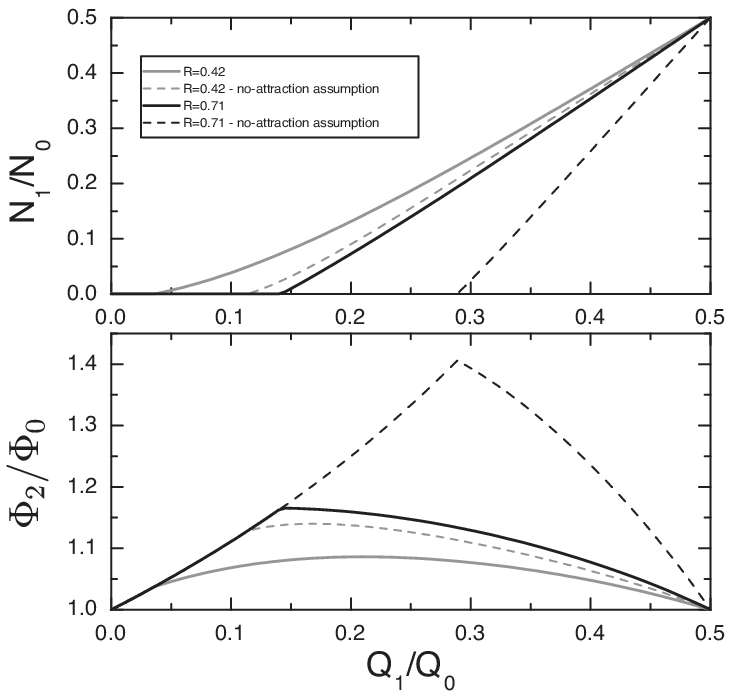}

\caption{Full lines:  spheres relative distribution $N_1/N_0$ and volume fraction $\Phi_2/\Phi_0$ as a function of flow rate distribution  $Q_1/Q_0$ from our two-dimensional simulations for two representative radii $R=0.42$ and $R=0.71$. The curves are straightforwardly derived from equations (\ref{eq:n1}) and \ref{eq:q1} and computed values of $y_0^*$ and $u_x^*$ (table \ref{table:quartic}). The results are compared with the hypothesis where the particles would follow the streamlines ($y_0^*=y_f$) (dashed lines).}\label{fig:volfrac}
\end{center}
\end{figure}

The evolution of $N_1/N_0$ as a function of $Q_1/Q_0$ for two-dimensional rigid spheres is shown on figure \ref{fig:volfrac} for two representative radii. By symmetry, considering $Q_1<Q_2$ is sufficient. In order to discuss the enrichment in particles in the high flow rate branch (branch 2 then), it is also convenient to consider directly the volume fraction variation $\Phi_2/ \Phi_0=(N_2/Q_2)/(N_0/Q_0)$.

When $Q_1/Q_0=1/2$, the particles flow splits equally into the two branches: $N_1=N_0$ and $\Phi_2=\Phi_0$. For all explored sizes of spheres, when the flow rate in branch 1 decreases, there is an enrichment in particles in branch 2, which is precisely the Zweifach-Fung effect: $N_1/N_0<Q_1/Q_0$ or $\Phi_2/\Phi_0>1$. Then, even in the most homogeneous case, the attraction towards the low flow rate branch is not strong enough to counterbalance the depletion effect that favours the high flow rate branch. However, this attraction effect cannot be considered as negligible, in particular for large particles: while, in case the particles follow their underlying fluid streamline, the maximum enrichment in the high flow rate branch  would be around $40\%$ for $R=0.71$, it drops down to less than $17 \%$ in reality. Similarly, the critical flow rate ratio $Q_1/Q_0$  below which no particle enters into branch 1 is greatly shifted: from around $0.29$ to around $0.15$ for $R=0.71$. For smaller spheres ($R=0.42$), this asymmetry in the distribution between the two branches is weak: while the maximum enrichment in the high flow rate branch would be around $15\%$ in a no-attraction case, it drops to less than $8\%$ due to the attraction towards the low flow rate branch.

When the flow $Q_1$ is equal to zero or $Q_0/2$, $\Phi_2$ is equal to $\Phi_0$; thus  there is a maximal enrichment for some flow rate ratio between 0 and 1/2. The increase in $\Phi_2$ with the decrease of $Q_1$ (right part of the curves of figure \ref{fig:volfrac}) is mainly due to the decrease of the relative importance of the free layer near the wall on the side of branch 2. Two mechanisms are responsible for the decrease of $\Phi_2$ when $Q_1$ decreases (left part of the curves of figure \ref{fig:volfrac}): first, when no particle can enter the low flow rate branch because its hypothetical separation line $y_0^*$ is above its maximum position $1-R$, then the high flow rate branch receives only additional solvent when $Q_1/Q_0$ decreases and its particles are more diluted. Then all curves fall down on the same curve  $\Phi_2/ \Phi_0=1/(1-Q_1/Q_0)$ corresponding to $N_1=0$ (or $N_2=N_0$), which results in a sharp variation as $Q_1/Q_0$ goes trough the critical flow rate ratio. A smoother mechanism is also to be taken into account here, as it is finally the one that determines the maximum for smaller $R$. As $Q_1/Q_0$ decreases, branch 2 recruits fluid and particles that are closer and closer to the opposite wall. As seen in figure \ref{fig:profils}, the discrepancy between the flow and particle velocities increases near the walls, so that $N_2$ increases less than $Q_2$: the resulting concentration in branch 2 finally decreases.

Finally, for applicative purposes, the consequences of the attraction towards the low flow rate branch are twofold: if one wishes to obtain a particle-free fluid (e.g. plasma without red blood cells), one has to set $Q_1$ low enough so that $N_1=0$. Due to attraction towards the low flow rate branch, this critical flow rate is decreased and the efficiency of the process is lowered. If one prefers to concentrate particles, then one must find the maximum of the $\Phi_2/\Phi_0$ curve. This maximum is lowered and shifted by the attraction towards the low flow rate branch (see figure \ref{fig:volfrac}). Note that for small spheres (e.g. $R=0.42$) the position of the maximum does not correspond to the point where $N_1$ vanishes; in addition, the shift direction of the maximum position depends on the spheres size: while it shifts to lower $Q_1/Q_0$ values for $R=0.71$, it shifts to higher values for $R=0.42$. 

The choice of the geometry, within our symmetric frame, can also greatly modify the efficiency of a device. Since the depletion effect eventually governs the final distribution, narrowing the inlet channel is the first requirement. On the other hand, it also increases the attraction towards the low flow rate branch, but one can try to diminish it. As discussed in the preceding section, this can be done by widening reasonably the daughter branches. For instance, if their half-width is not 1 but 2.5, as in figure \ref{fig:traj}(b), the  slope $\xi$ in the law $y_0^*= \xi \times(Q_1/Q_0-1/2)$ increases by around $15\%$ for $R=0.67$. The  critical $Q_1/Q_0$ below which no particle enters the low flow rate branch increases from 0.13 to 0.19, which is good for fluid-particle separation, and the maximum enrichment $\Phi_2/\Phi_0$ that can be reached is $22\%$ instead of $15\%$. Alternatively, since the attraction is higher in two dimensions than in three, we can also infer that considering thicker channels, which does not modify the depletion effect, can greatly improve the final result. Note that this conclusion would have been completely different in case of high flow rate branch enrichment due to some attraction towards it, as claimed in some papers: in such a case, confining as much as possible would have been required, as it increases all kinds of cross-streamline drifts.

\subsection{Consistency with the literature}
\label{sec:consistency}

\begin{figure}
\begin{center}

\includegraphics[width=\columnwidth]{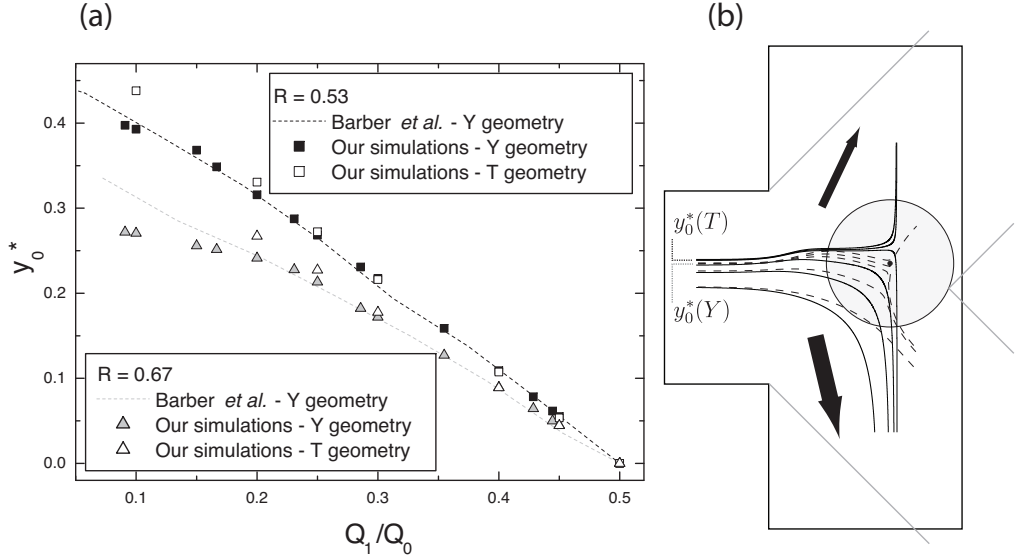}

\caption{(a) Position of the particle separation line $y_0^*$  in a symmetric Y-shaped bifurcation: according to \cite{barber08} (data extracted from figure 4 of the cited paper) and according to our simulations. The results for similar spheres in our T geometry are also shown. (b) Trajectories from our simulations in the T- and Y- shaped bifurcation, for similar sphere size ($R=0.67$) and flow rate ratio ($Q_1/Q_0=0.2$). Full lines: T geometry; dashed lines: Y geometry. The corresponding separating streamline positions (respectively, $y_0^*(T)$ and $y_0^*(Y)$) are also indicated. The sphere that is depicted is located at its stagnation point $y_0'$ (see figure \ref{fig:traj}) in the Y geometry.}\label{fig:conclulitt}
\end{center}
\end{figure}

\begin{figure}
\begin{center}

 \includegraphics[width=\columnwidth]{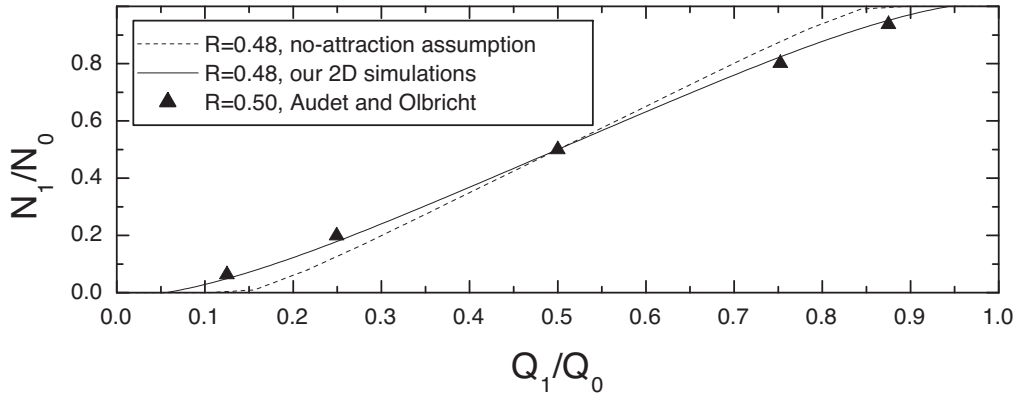}

\caption{Particle distribution as a function of flow rate ratios for spheres of medium size, according to our simulations and according to the simulations shown in \cite{audet87} (same data as plotted in figure \ref{fig:compar}).}\label{fig:concluAudet}
\end{center}
\end{figure}

We now come back to the previous studies already discussed in section \ref{sec:litt} in order to check the consistency between them and our results. 

The only paper dealing with the position of the particle separating streamline was the one by \cite{barber08}, where a symmetric Y-shaped bifurcation is studied (branches leaving the bifurcation with a $45^\circ$ angle relatively to the inlet channel, see figure \ref{fig:schema}(a)). In figure \ref{fig:conclulitt}(a) we compare their results with our simulations in a similar geometry. The agreement between the two simulations (based on two different methods) is very good, except for large particles ($R=0.67$) and low $Q_1/Q_0$. Note that Barber {\itshape et al.} have chosen to consider branches whose widths follow the law $w_0^3=w_1^3+w_2^3$, where $w_0$ is the width of the inlet branch and $w_1$ and $w_2$ are the widths of the daughter branches. This law has been shown to describe approximately the relationship between vessel diameters in the arteriolar network \cite[see][]{mayrovitz83}. With our notations, they thus consider $a=\sqrt[3]{1/2}\simeq 0.79$, while we focused on $a=1$ in order to compare with the T-shaped bifurcation. In addition, their apex has a radius 0.75 (for the $R=0.67$ case) while ours is sharper (radius of 0.1). These differences seem to impact only partly the results, as discussed above. We can expect this slight discrepancy to be due to the treatment of the numerical singularities that appear when the particle is close to one wall. For $R=0.67$, the maximum position $y_0$ is 0.33, which is close to the separating streamline position.

It is also interesting to compare our results in the Y-shaped bifurcation with the results in the T geometry, which was chosen to make the discussion easier. We can see that, for low enough $Q_1/Q_0$, the attraction towards the low flow rate branch is slightly higher. This can be understood by considering a particle with initial position $y_0$ slightly below the critical position $y_0^*$ found in the T geometry: in the latter geometry, it will eventually enter the high flow rate branch, by definition of $y_0^*$. As shown in figure \ref{fig:conclulitt}(b), in the Y geometry, this movement is hindered by the apex since the final attraction towards the high flow rate branch occurs near the opposite wall (the second effect discussed in section \ref{sec:discussion}). Finally from this comparison we see that comparing results in T and symmetric Y geometry is relevant but for highly asymmetric flow distributions.\\

In section \ref{sec:litt}, the analysis  of the  two-dimensional simulations for $R=0.5$ spheres shown in \cite{audet87} showed  that there should be some attraction towards the low flow rate branch. Our simulations for $R=0.48$ showed that this effect is non negligible (figure \ref{fig:lignessep}b) and modifies greatly the final distribution (figure \ref{fig:volfrac}). Finally, we can see in figure \ref{fig:concluAudet} that our simulations give similar results as the simulation by Audet and Olbricht.

As for the experiments presented in \cite{yang06} for $R=0.46$, we showed that the final distribution was consistent with a no-attraction assumption. As we showed in figure \ref{fig:lignessep}(a), in a three-dimensional case, the attraction towards the low flow rate region is weak for spheres of radius $R\simeq 0.5$ or smaller, which is again coherent with the results of Yang \textit{et al.}. Note that, while their results were considered by the authors as a basis to discuss some attraction effect towards the high flow rate branch, we see that their final distributions are just reminiscences of the depletion effect in the inlet channel.

The other consistent set of studies in the literature deals with  large balls in three dimensional channels. We have studied balls of radius $R=0.71$ that stop entering branch 1 when $Q_1/Q_0\lesssim 0.22$ (figure \ref{fig:lignessep}a), while this critical flow rate would be around $0.29$ in case they would follow the fluid streamlines. This critical flow rate is expected to be slightly higher for larger balls of radius $R\simeq0.8$, but far lower than $0.35$, which would be the no-attraction case. In the experiments of \cite{roberts06}, some balls are still observed in branch 1 when $Q_1/Q_0\simeq 0.22$ (figure \ref{fig:compar}), indicating a stronger attraction effect towards the low flow rate branch, which can be associated to the fact that the authors considered a square cross section channel, while the confinement in the third direction is $0.5 < 0. 71$ in our case. The experiments with circular cross section channels lead to contradictory results: in \cite{chien85} and \cite{ditchfield96}, the results were  consistent with a no-attraction assumption, therefore they are in contradiction with our results. On the contrary, in \cite{roberts03}, the critical flow rate for $R=0.77$ is around 0.2, which would show a stronger attraction than in our case. Note that all these apparently contradictory observations are to be considered keeping in mind that the data of $N_1/N_0$ as a function of $Q_1/Q_0$ are sometimes very noisy in the cited papers.

\section{Conclusion}

In this paper, we have focused explicitly on the existence and direction of some cross streamline drift of particles in the vicinity of a bifurcation with different flow rates in the daughter branches. A new analysis of some previous unexploited results of the literature first gave us some indications on the possibility of an attraction towards the low flow rate branch.

Then the first direct experimental proof of attraction towards the low flow rate branch was shown and arguments for this attraction were given with the help of two-dimensional simulations. In particular, we showed that this attraction is the result of two antagonistic effects: the first one, that takes place at the entrance of the bifurcation, induces migration towards the low flow rate branch, while the second one takes place near the stagnation point and induces migration towards the high flow rate branch but is not strong enough, in standard configurations of branches of comparable sizes, to counterbalance the first effect.

This second effect is the only one that was previously considered in most papers of the literature, which has lead to the misleading idea that the enrichment in particles in the high flow rate branch is due to some attraction towards it. On the contrary, it had been argued by Barber \textit{et al.} that there should be some attraction towards the low flow rate branch. By distinguishing the two effects mentioned above, we have tried to clarify their statements.

In a second step, we have discussed the consequences of such an attraction on the final distribution of particles. It appears that the attraction is not strong enough, even in a two-dimensional system where it is stronger, to counterbalance the impact of the depletion effect. Even in the most homogeneous case where the particles are equally distributed across the channel but cannot approach the wall closer than their radius, the existence of a free layer near the walls favours the high flow rate branch, which eventually receives more particles than fluid.

However, these two antagonistic phenomena are of comparable importance, and none can be neglected: the particle volume fraction increase in the high flow rate branch is typically divided by two because of the attraction effect. On the other hand, the initial distribution is a key parameter for the prediction of the final splitting.  For deformable particles, initial lateral migration can induce a narrowing of their distribution, which will eventually favours the high flow rate branch. For instance, in \cite{barber08}, the authors had to adjust the free layer width in their simulations in order to fit experimental data on blood flow. On the other hand, in a network of bifurcations, the initially centered particles will find themselves close to one wall after the first bifurcation, which can favour a low flow rate branch in a second bifurcation.

Note finally that, as seen in \cite{enden92}, these effects become weaker when the confinement decreases. Typically, as soon as the sphere diameter is less than half the channel width, the variations of volume fraction do not exceed a few percent. 

For applicative purposes, the consequences of this attraction have been discussed and some prescriptions have been proposed. Of course, one can go further than our symmetric case and modify the angle between the branches, or consider many-branch bifurcations, and so on.  However, the T-bifurcation case allowed to distinguish between two goals: concentrating a population of particles, or obtaining a particle-free fluid. The optimal configuration can be different according to the chosen goal. Similar considerations are also valid when it is about doing some sorting in polydisperse suspensions, which is an important activity  \cite[see][]{pamme07}: getting an optimally  concentrated suspension of big particles might not be compatible with getting a suspension of small particles free of big particles.\\

Now that the case of spherical particles in a symmetric bifurcation has been studied and the framework well established, we believe that quantitative discussions could be made in the future about the other parameters that we put aside here. In particular, discussing the effect of the deformability of the particles is a challenging problem if one only considers the final distribution data, as the deformability modifies the initial distribution, but most probably also the attraction effect. In a network, the importance of these contributions will be different according, in particular, to the distance between two bifurcations, so they must be discussed separately.

Considering concentrated suspensions is of course the next challenging issue. Particles close to each other will obviously hydrodynamicaly interact, but so will distant particles, through the modification of the effective resistance to flow of the branches. In such a situation, considering pressure driven or flow rate driven fluids will be different.

For concentrated suspensions of deformable particles in a network, like blood in the circulatory system, the relevance of a particle-based approach can be questioned. Historical models for the major blood flow phenomena are continuum models with some ad-hoc parameters, which must be somehow related to the intrinsic mechanical properties of the blood cells (for a recent example, see \cite{obrist10}). Building up a bottom-up approach in such a system is a long quest. For dilute suspensions, some links between the microscopic dynamics of lipid vesicles and the rheology of a suspension have been recently established \cite[see][]{danker07,vitkova08,ghigliotti10}. For red blood cells, that exhibit qualitatively similar dynamics \cite[see][]{abkarian07,deschamps09,noguchi10,farutin10,dupire10}, we can hope that such a link will soon be established, following \cite{vitkova08}. For confined and concentrated suspensions, the distribution is known to be non homogeneous, which has direct consequences on the rheology (the Fahraeus-Lindquist effect). Once again, while empirical macroscopic models are able to describe this reality, establishing the link between the viscosity of the suspension and the local dynamics is still a challenging issue. The final distribution of the flowing bodies is the product of a balance between migration towards the center, which has already been discussed in the introduction of the present paper, and interactions between them that can broaden the distribution \cite[see][]{kantsler08,podgorski10}. The presence of deformable boundaries also needs to be taken into account, as shown in \cite{beaucourt04}. In the meantime, the development of simulations techniques for quantitative three-dimensional approaches  is a crucial task, which is becoming more and more feasible \cite[see][]{McWhirter09,biben10}.

\begin{acknowledgments}
The authors thank G. Ghigliotti for his final reading and acknowledge financial support from ANR MOSICOB and from CNES.
\end{acknowledgments}

\oneappendix

\section{}\label{ap:numeric}

In this appendix, details for the derivation of   equations (\ref{eq:3bs})-(\ref{eq:4bs})-(\ref{eq:5bs})  from equations (\ref{eq:1b})-(\ref{eq:2b})-(\ref{eq:2ab})  are given.

We introduce first the scalar product in $L^2(\Omega_f)^2$ as follows:
$$
    \forall \f,\g\in L^2(\Omega_f)^2,\quad
    <\f,\g>_{L^2(\Omega_f)^2}=\int_{\Omega_f}\f\bcdot\g.
$$
The variational formulation of problem
(\ref{eq:1b})-(\ref{eq:2b})-(\ref{eq:2ab})
is obtained by taking the scalar product
of the equation (\ref{eq:1b}) in $L^2(\Omega_f)^2$
with a test
function $\v\in H^1_0(\Omega_f)^2$ and we multiply equation
(\ref{eq:2b}) by a test function $q\in L^2_0(\Omega)$. It leads to this
problem: { find 
$(\u,p)\in H^1(\Omega_f)^2\times L^2_0(\Omega_f)$ such that:}
\begin{eqnarray}
  \label{eq:13}
  -2\nu\int_{\Omega_f}(\bnabla\bcdot\Tau(\u))\bcdot\v + \int_{\Omega_f}\bnabla
  p\bcdot\v &=& 0, \forall \v\in H^1_0(\Omega_f)^2,\\
  \label{eq:13a}
  \int_{\Omega_f}q\bnabla\bcdot\u &=& 0,\forall q\in
  L^2_0(\Omega_f),\\
  \label{eq:13b}
  \u &=& \f \mbox{ on } \partial\Omega_f.
\end{eqnarray}
Applying Green's formula to equation (\ref{eq:13}) we obtain
\begin{eqnarray}
  \nonumber
  2\nu\int_{\Omega_f}\Tau(\u):\bnabla\v 
  &-& 2\nu\int_{\partial\Omega_f}\Tau(\u)\n\bcdot\v \\
  \label{eq:14}
  &-&\int_{\Omega_f}p\bnabla\bcdot\v 
  +\int_{\partial\Omega_f}p\v\bcdot\n = 0.
\end{eqnarray}
Where $\n$ denotes the outer unit normal on $\partial\Omega_f$.
Taking into account that $\v$ vanishes on
$\partial\Omega_f$ (recall that we have chosen the test function
$\v\in H^1_0(\Omega_f)^2$), the problem
(\ref{eq:13})-(\ref{eq:13a})-(\ref{eq:13b}) is now equivalent to this
one: { find 
$(\u,p)\in H^1(\Omega_f)^2\times L^2_0(\Omega_f)$ such that:}
\begin{eqnarray}
  \label{eq:3}
  2\nu\int_{\Omega_f}\Tau(\u):\bnabla\v 
  -\int_{\Omega_f}p\bnabla\bcdot\v 
  &=& 0, \forall
  \v\in H^1_0(\Omega_f)^2, \\
  \label{eq:4}
  \int_{\Omega_f}q\bnabla\bcdot\u &=& 0,
  \forall q\in L^2_0(\Omega_f),\\
  \label{eq:5}
  \u &=& \f \mbox{ on } \partial\Omega_f.
\end{eqnarray}
Note that $\Tau(\u)$ is symmetric
($\Tau(\u):\bnabla\v = \Tau(\u):(\bnabla\v)^t$). So that we can write
$\Tau(\u):\bnabla\v = \Tau(\u):\Tau(\v)$. Finally, the variational
formulation of our initial problem
(\ref{eq:1})-(\ref{eq:2})-(\ref{eq:2a}) is given by:
{find 
$(\u,p)\in H^1(\Omega_f)^2\times L^2_0(\Omega_f)$ such that:}
\begin{eqnarray}
  \label{eq:3b}
  2\nu\int_{\Omega_f}\Tau(\u):\Tau(\v)
  -\int_{\Omega_f}p\bnabla\bcdot\v 
  &=& 0, \forall
  \v\in H^1_0(\Omega_f)^2, \\
  \label{eq:4b}
  \int_{\Omega_f}q\bnabla\bcdot\u &=& 0,
  \forall q\in L^2_0(\Omega_f),\\
  \label{eq:5b}
  \u &=& \f \mbox{ on } \partial\Omega_f.
\end{eqnarray}

\begin{remark}
  As we have
  \begin{equation}
    \label{eq:9}
    \Tau(\u):\Tau(\v) =
    \Tau(\u):\bnabla\v
    =\frac{1}{2}\bnabla\u:\bnabla\v +
    \frac{1}{2}(\bnabla\u)^t:\bnabla\v,
  \end{equation}
the first integral in equation (\ref{eq:3b}) can be rewritten thanks
to this identity 
\begin{equation}
  \label{eq:10}
  \int_{\Omega_f} \Tau(\u):\Tau(\v) =
  \frac{1}{2} \int_{\Omega_f}\bnabla\u:\bnabla\v.
\end{equation}
Indeed,  by integration by part and using the incompressibility
constraint $\bnabla\bcdot\u=0$ we have
${\displaystyle \int_{\Omega_f}(\bnabla\u)^t:\bnabla\v = 0}$. Thus we
can retrieve the formulation of our problem as a minimization of a
kind of energy. The velocity field $\u$ is then the solution of this problem
\begin{equation}
  \label{eq:6}
  \mathbf{J}(\u) =
  \inf_{
    \stackrel{
      \v\in H^1(\Omega_f)^2
    }
    {
      \bnabla\bcdot\v = 0, \v_{|\partial\Omega_f}=\f
    }
  }
  \mathbf{J}(\v),
\end{equation}
where
\begin{equation}
  \label{eq:7}
  \mathbf{J}(\v) =
  \frac{1}{2}  \nu\int_{\Omega_f}\bnabla\v:\bnabla\v
   = \nu\int_{\Omega_f}\Tau(\v):\Tau(\v).
\end{equation}
\end{remark}

\end{document}